\documentclass[journal,comsoc]{IEEEtran}

\usepackage[T1]{fontenc}

\ifCLASSINFOpdf

\else

\fi

\usepackage{amsmath}

\usepackage[cmintegrals]{newtxmath}

\usepackage{pifont} 		    
\usepackage{threeparttable}     
\usepackage{multirow}        	
\usepackage{tabularx}       	
\usepackage{graphicx}           
\usepackage{subfigure}          
\usepackage{booktabs}			
\usepackage{hyperref}           
\usepackage{color}              
\usepackage{hyperref}           
\usepackage{enumerate}          
\usepackage[switch]{lineno}     
\usepackage{ulem}               

\hypersetup{
colorlinks=true,
linkcolor=black
}

\begin{document}

\title{An Anti-disguise Authentication System Using the First Impression of Avatar in Metaverse}

\author{
Zhenyong~Zhang,
Kedi~Yang,
Youliang~Tian,
and Jianfeng~Ma

\thanks{Manuscript received XX, 2023; revised XX, 2024; accepted XX, 2024. Date of publication XX, 2024; date of current version XX, 2024. This work is supported by the National Key Research and Development Program of China under Grant 2021YFB3101100; National Natural Science Foundation of China under Grant No. 62362008, No. 62303126,  No.62262058, No. 62272123;  Guizhou Provincial Science and Technology Projects No. ZK[2022]149; Project of High-level Innovative Talents of Guizhou Province under Grant [2020]6008; Science and Technology Program of Guizhou Province under Grant [2020]5017, [2022]065; Guizhou Provincial Research Project (Youth) for Universities under grant [2022]104; Guizhou Provincial Postgraduate Research Fund under Grant No.YJSKYJJ[2021]029; supported by the open project of the Key Laboratory of Computing Power Network and Information Security, Ministry of Education. The associate editor coordinating the review of this manuscript and approving it for publication was Erisa Karafili. (\textit{Zhenyong Zhang and Kedi Yang contributed equally to this work.) (Corresponding author: Youliang Tian)}}
\thanks{Zhenyong Zhang is with the State Key Laboratory of Public Big Data, College of Computer Science and Technology, 
Guizhou University, Guiyang 550025, China, and also with the Key Laboratory of Computing Power Network and Information Security, Ministry of Education, Qilu University of Technology (Shandong Academy of Sciences), Jinan, Shandong 250353, China (e-mail: zyzhangnew@gmail.com)

Kedi Yang and Youliang Tian  are with the State Key Laboratory of Public Big Data, College of Computer Science and Technology, 
Guizhou University, Guiyang 550025, China, and also with the Institute of Cryptography \& Data Security, GuiZhou University, Guiyang 550025, China  (e-mail: kdyang.gz@gmail.com; youliangtian@163.com).
}

\thanks{Jianfeng Ma is with the School of Cyber Engineering, Xidian University,
Xi’an 710126, China, and also with the State Key Laboratory of Public Big
Data, College of Computer Science and Technology, Guizhou University,
Guiyang 550025, China (e-mail: jfma@mail.xidian.edu.cn).}

\thanks{
Digital Object Identifier XXX
}
}

\markboth{Submitted to IEEE Trans, 2023}
{Shell \MakeLowercase{\textit{et al.}}: Bare Demo of IEEEtran.cls for IEEE Communications Society Journals}

\maketitle

\begin{abstract}
Metaverse is a vast virtual world parallel to the physical world, where the user acts as an avatar to enjoy various services that break through the temporal and spatial limitations of the physical world. Metaverse allows users to create arbitrary digital appearances as their own avatars by which an adversary may disguise his/her avatar to fraud others. In this paper, we propose an anti-disguise authentication method that draws on the idea of the first impression from the physical world to recognize an old friend. Specifically, the first meeting scenario in the metaverse is stored and recalled to help the authentication between avatars. To prevent the adversary from replacing and forging the first impression, we construct a chameleon-based signcryption mechanism and design a ciphertext authentication protocol to ensure the public verifiability of encrypted identities. The security analysis shows that the proposed signcryption mechanism meets not only the security requirement but also the public verifiability. Besides, the ciphertext authentication protocol has the capability of defending against the replacing and forging attacks on the first impression. Extensive experiments show that the proposed avatar authentication system is able to achieve anti-disguise authentication at a low storage consumption on the blockchain.

\end{abstract}


\IEEEpeerreviewmaketitle

\section{Introduction}

\IEEEPARstart{M}{etaverse} is an immersive virtual environment simulating and extending the physical world \cite{Wang2023Survey, Wang2023Metaverse}. People live in the metaverse acting as any object as they want to enjoy the digital life.

In the metaverse, the user creates their own digital actor, termed avatar, to be the identity in the virtual world \cite{Genay2022RealAvatar}, which can be a strange animal or a human-like model \cite{Shen2023XAvatar, Ho2023Editable}. With the development of virtual reality (VR) and artificial intelligence (AI), the metaverse becomes the highly-qualified second living space of human beings’ coexistence with the physical world \cite{Duan2021Good,Cheng2022NextG}. In the social case, people can launch meetings with others through avatars and perceive the micro-expressions to achieve immersive chat. In the business case, the staff guides consumers through avatars to realize immersive shopping. 

Although the metaverse provides us with immersive services \cite{Wang2022Edu,Jiang2023Game,Tsai2023medicine} the disguise attack is a potential threat that the adversary creates and mimics someone's avatars to fraud other users and steal their private information  \cite{Soliman2023DigtHuman}. The reason why the disguise attack is possible is that the user tends to trust the avatar with the familiar appearance and voice while is not willing to think that the ``friend" is fake. A company in Hong Kong reported that it was deceived more than HK\$200m (£20m) because an employee received a deepfake video conference call\footnote{\href{https://www.theguardian.com/world/2024/feb/05/hong-kong-company-deepfake-video-conference-call-scam}{www.theguardian.com}}. During the conference, the adversary disguised the senior officer of the company, who looked like the true one, thereby deceiving the employee into transferring funds to the designated bank account. The security and privacy issues of the metaverse have been deeply analyzed with a high-level perspective in \cite{Falchuk2018Privacy}.

What's worse is that the current metaverse application is affected by the performance of VR hardware and uses a simple digital appearance as the user's avatar \cite{Aseeri2021Influence}, which provides convenience for the adversary to disguise the appearance. For example,  \textit{Roblox}, the first listed company of metaverse adopts Lego brick man as its avatar, while, \textit{Meta}, the formerly Facebook uses a floating torso as its avatar, which allows adversaries to create the same avatar easily. Using the deepfake technique, AI-driven avatars can be the same as human-driven avatars in appearance and sound, which poses a huge challenge to retain a safe social environment in the metaverse.

\subsection{An Illustration Of Disguise Attack}
The disguise attack is designed and executed based on social engineering. Wang \textit{et al.} \cite{Wang2023Survey} described how an attacker uses the digital replica to construct a fake avatar, which can deceive, fraud, and even commit crimes against the victim’s friends in the metaverse.

\begin{figure}[htbp]
\begin{center}
    \includegraphics[width=0.35\textwidth]{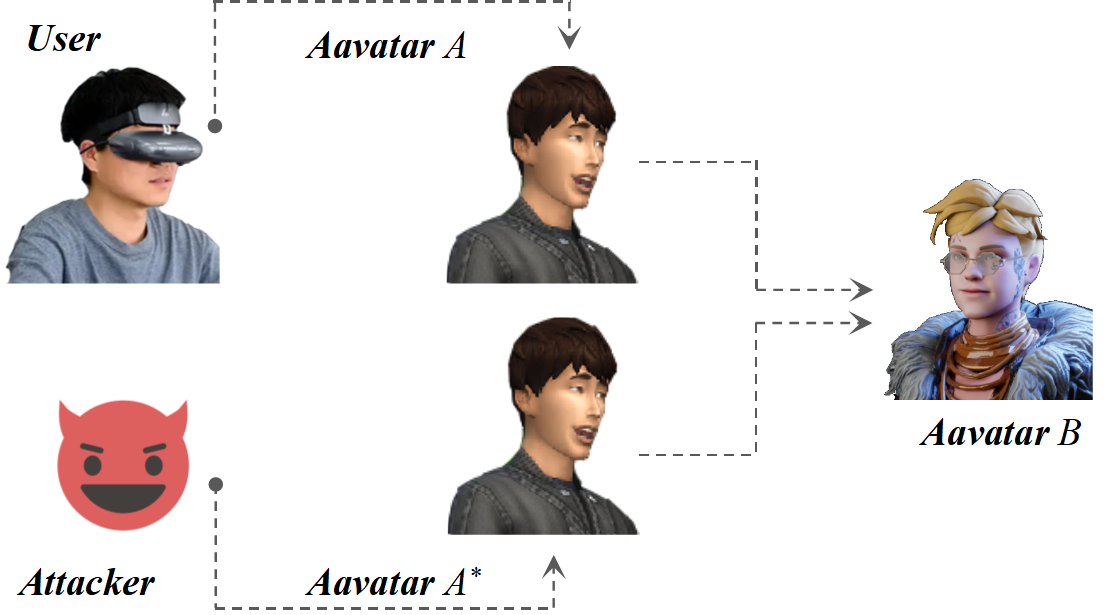}
    \caption{\small{An illustration of the disguise attack, where an adversary generates the same avatar $A^*$ as avatar $A$ to deceive avatar $B$.}}
    \vspace{-1.0em}
    \label{fig:DiguiseAttack}
 \end{center}
\end{figure}

To evaluate the impact of the disguise attack, we build an environment by combining a voice plugin and two metaverse platforms \textit{Xirang} and \textit{VS Work}. During the attack, the adversary trains a voice model of the victim based on the voice plugin and then talks with ten friends of the victim using the voice model. The attack has two goals. One is to deceive the friends to disclose his/her private information such as ID card number. The other is to induce the friends to perform designated actions. The successful rate of the attack is shown in TABLE \ref{tab:DiguiseAttack_0}.

We can see that the adversary on the \textit{Xirang} platform can successfully obtain personal information with a probability of 50\%, while the probability of inducing friends to perform designated actions is 40\%. On the platform of \textit{VS Work}, the probability that the adversary successfully obtains the private information and induces the friends to perform designated actions is 60\% and 30\%, respectively. From the above results, it is easy for the adversary to execute the disguise attack in the metaverse platform.

\begin{table}[htbp]
    \footnotesize
      \centering
      \caption{Successful rate of the disguise attack in the metaverse platform}
      \label{tab:DiguiseAttack_0}
      \begin{threeparttable}
      \begin{tabular}{p{40 pt}<{\centering}p{30 pt}<{\centering}p{60 pt}<{\centering}p{60 pt}<{\centering}}
          \hline
        \specialrule{0em}{1pt}{1pt}
          Metaverse platform &  Display &  Rate of providing private information &  Rate of performing designated actions  \\
        \specialrule{0em}{1pt}{1pt}
           \hline
          \specialrule{0em}{1pt}{1pt}
            $Xirang$  & $/$ & 50\% & 40\% \\
            $VS\; Work$  & Name & 60\% & 30\%  \\
      \specialrule{0em}{1pt}{1pt}
        \hline
      \end{tabular}
       \begin{tablenotes}
        \footnotesize
         \item[] ``$/$'' indicates that the platform does not display any identity information on the interaction screen.
      \end{tablenotes}
\end{threeparttable}
\vspace{-0.5cm}
\end{table}

\subsection{Traditional Authentication Approaches}
Traditional authentication systems combine multi-factor identities such as cryptographic keys, biometric features, and account-password to realize the login authentication. In metaverse, however, entering the account and password is not user-friendly, rendering its application for the authentication. In the following, we compare the well-known authentication methods and present our idea.

\subsubsection{Cryptographic Keys} 
The first idea is to leverage cryptographic keys such as a pre-shared symmetric key \cite{Gope2019LightTwo} and public key \cite{Shang2020Secure}. The key is used as an identity factor of the avatar and integrated into the challenge-response mechanism to verify the avatar's identity. With the pre-shared symmetric key, one avatar acting as a verifier throws a random challenge to the friend acting as a prover who returns a ciphertext associated with the challenge based on the shared symmetric key. If the ciphertext can be decrypted correctly, the verification process passes; otherwise, the verification process fails. With this method, however, if the adversary and the victim's friend are both friends, the adversary can pass the verification process because he/she has a pre-shared symmetric key with the victim's friends, enabling the adversary to generate a correct response based on the symmetric key. Therefore, the authentication method that leverages a pre-shared symmetric key as an identity factor of the avatar cannot defend against the disguise attack executed by the acquaintance.

The public-key approach is similar to the symmetric-key method. It uses a cryptographic challenge-response protocol to authenticate the avatars  \cite{Ryu2022MutMeta}. With this approach, the verifier gets a public key from the prover and throws a random challenge to the prover who then generates a response. Once the received response matches the prover's public key, the verifier regards the prover as a trusted friend even if the prover is a disguised person. This is because the public key and response are both provided by the adversary. Therefore, the approach using the public key as the identity factor of the avatar fails to prevent the disguise attack.

\subsubsection{Biometric features} The second idea is to incorporate various biometric features of the manipulator into his/her avatar as identity factors such as face and voice \cite{Ryu2022MutMeta}. Since the head-mounted display (HMD) completely covers the eyes, the iris is more suitable for authentication in the metaverse compared to other biometric features. During the authentication, the verifier obtains the biometric template from the prover and stimulates the prover to generate a biometric sample. As long as the sample matches the template, the authentication is passed. This method encounters the same vulnerability as the public key method, that is, both the biometric template and sample are provided by the adversary. Therefore, incorporating biometric features into avatars as identity factors fails to prevent the disguise attack in the metaverse.

\subsubsection{Display Identifier} 
A simple way to authenticate is to display the user's ID on the screen. However, this method brings a great burden for users because they need to remember the complicated ID numbers. For example, a friend's ID number is $ID=5700121517963$ and an adversary's ID number is  $ID^\prime=5700121511963$. It is highly possible for the user to wrongly recognize that these two ID numbers are the same.

\subsection{Our Idea And Contributions} 
In the real world, people recognize their friends primarily based on their own abilities. For example, when a person decides whether a stranger is a friend or not, he/she usually relies on whether they can recall a meeting scene with the friend or not. The first meeting is often used as a common and precious memory between two friends. People have a deep perception of the first impression of a stranger, which lasts a long time in their memory \cite{Willis2006FI}. No matter how the friend's appearance has changed, the unconscious brain activity helps people recognize the friend \cite{Waroquier2010FI}.

Inspired by this human perception, we propose an anti-disguise authentication method based on the first impression to identify avatars in the metaverse. We combine the avatar's core identity with the ``first impression" containing the avatar's appearance and the metaverse circumstance. By integrating the first impression into the authentication protocol, we provide the verifier with an auxiliary authentication factor that is verified using the verifier's own recall. Based on this factor, the user can quickly recognize whether the friend is legitimate or not.

\begin{figure}[htbp]
\begin{center}
    \includegraphics[width=0.35\textwidth]{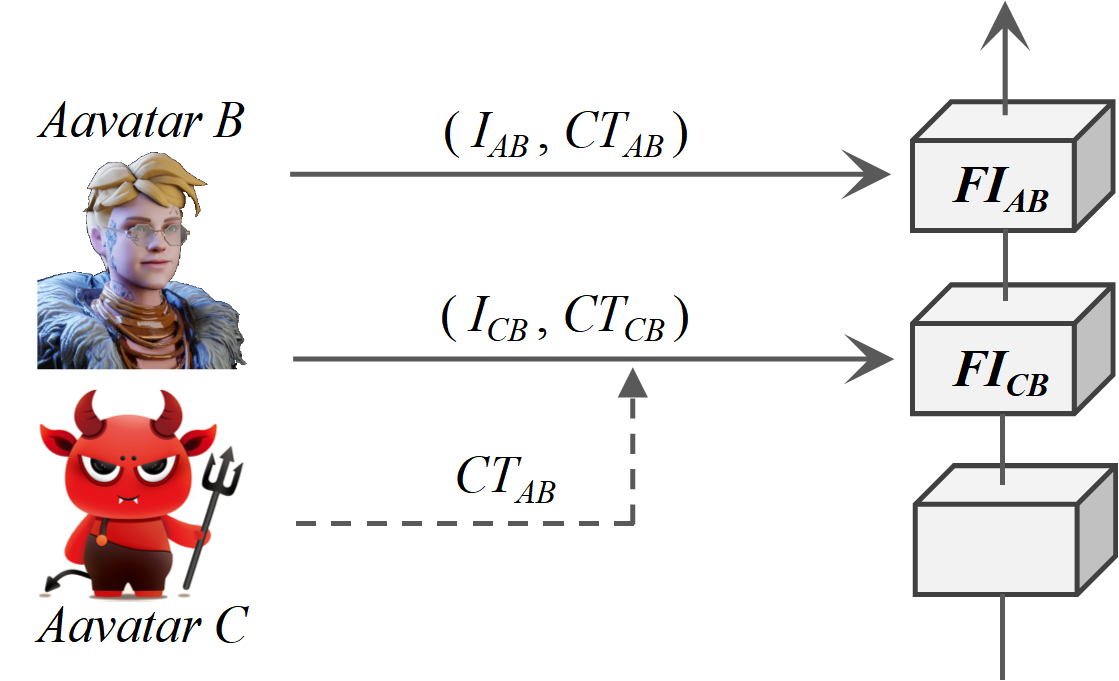}
    \caption{\small{The replacing attack on the first impression. The notation $CT_{AB}$ represents the ciphertext of the first impression that is made by $A$ for $B$ and $CT_{CB}$ represents the ciphertext of the first impression that is made by  $C$ for $B$.}}
      \label{fig:Challenge}
 \end{center}
\end{figure}

Traditional public-key encryption such as RSA \cite{Rivest1978RSA} is a practical technique for achieving confidentiality of the first impression. However, the encrypted first impression is vulnerable to the replacing attack. An example is given in Fig.\ref{fig:Challenge}. The avatar $B$ writes the ciphertext $CT_{AB}$ containing the first impression $FI_{AB}=(I_{AB}, CT_{AB})$ of avatar $A$  into blockchain, where $I_{AB}$ is the extraction index. When $B$ writes the first impression of avatar $C$, the $C$ acting as an adversary replaces $CT_{CB}$ with $CT_{AB}$ to form $FI_{CB}=(I_{CB},CT_{AB})$. Thus, it misleads $B$ to treat $C$ as $A$ because the first impression is replaced. Therefore, the adversary $C$ successfully disguises as the avatar $A$ to get the trust from $B$.

Considering the replacing attack, the adversary exploits the vulnerability that the blockchain does not verify the identities implied in the ciphertext. Although the two elements $I_{*}$ and $CT_ {*}$ of the first impression are not consistent, they are regarded as normal. If a ciphertext authentication protocol is introduced into the storage on the blockchain, the replacing attack can be avoided and the disguise attack can be further defended. Signcryption \cite{Zheng1997Signcryption} is a typical signature-then-encryption technique that can achieve the verification of the sender's and receiver's identities. However, the traditional signcryption mechanism  \cite{Xiong2021Test, Chen2022Offline, Dohare2022CLASS}  needs to decrypt the ciphertext to verify its identity, which exposes the plaintext information and is not suitable for the highly secret scenario storing the first impressions. Therefore, it is urgent to construct a signcryption mechanism with public variability to verify the identity of ciphertext without decryption, protecting the identity factor of the first impression from being replaced. 

Chameleon signature is a mutable signature mechanism \cite{Krawczyk2000Chameleon, Mohassel2010One, Chen2014Chameleon}. If one holds the chameleon private key, the signature mechanism allows one to generate a new signature associated with the original one. Making use of this advantage, we propose a chameleon signcryption to realize public verifiability by modifying parameters related to the original plaintext. Based on the chameleon signcryption, we design a ciphertext authentication protocol to avoid the replacing attack on the first impression, which further defends against the disguise attack.

To sum up, this paper presents an anti-disguise authentication scheme for avatars based on the idea of the first impression. The main contributions are as follows:

\begin{itemize}

\item We propose a chameleon signcryption mechanism based on the chameleon collision signature,  which verifies the identities implied in ciphertext without decryption.

\item We design a ciphertext authentication protocol based on the chameleon signcryption, which defends against the replacing attack on the first impression.

\item We develop an avatar authentication protocol based on the first impression, which enhances the detectability of the avatar's identity.

\item We build an anti-disguise authentication system for users to create avatars and enjoy the metaverse services safely. The authentication system utilizes the inter-planetary file system (IPFS) to alleviate the storage burden of the first impression on blockchains.

\end{itemize}

The remainder of this paper is organized as follows. Some preliminaries are presented in Section \ref{section:pre}. The chameleon signcryption mechanism is given in Section \ref{sec:Signcryption}. Section \ref{sec:Framework} introduces the anti-disguise authentication framework. The details of the anti-disguise authentication protocol are provided in Section \ref{sec:Protocol}. The security analysis and the performance evaluation are given in Section \ref{sec:SecuAnaly} and Section \ref{sec:PerfEval},  respectively. The related works are presented in Section \ref{sec:RelaWork}. Finally, section \ref{sec:Conclu} concludes the paper.

\section{Preliminaries}\label{section:pre}
The purpose of this work is to construct an anti-disguise authentication system using the first impression. The public verifiability of a signcryption mechanism is the key to guaranteeing that the ciphertext in the pair of the first impression cannot be replaced by the adversary. In this section, we first present the metaverse authentication framework, then review the chameleon collision signature and the traditional signcryption.

\subsection{ Metaverse Authentication Framework}
The existing metaverse platforms such as \textit{Horizon Worlds} and \textit{Xirang} mainly utilize the method of account-password to achieve login authentication. The related studies about metaverse authentication combine multi-factor identities \cite{Ryu2022MutMeta,Yang2023Trace} such as account-password, cryptographic keys, and biometric features, to realize the login authentication and mutual authentication between avatars. 

\subsubsection{Login authentication} 
In \cite{Ryu2022MutMeta}, the users' VR device is treated as a trusted entity to store the fingerprint template and private key, which generates signatures based on elliptic curves. During the login authentication, the user first enters the account-password and then the VR device captures the user's fingerprint samples to compare with the fingerprint template. If the sample matches the template in the VR device, the device submits the user's pseudo-random identity and a signature corresponding to the current timestamp to the server. If these parameters pass the verification on the server, the server decides that the user is legitimate.

\subsubsection{Mutual authentication} 
The avatar authentication framework  \cite{Yang2023Trace} leverages iris features, chameleon collision signature, and blockchain to achieve decentralized mutual authentication and malicious avatar traceability. During the mutual authentication, the avatar $A$ acting as a prover provides his/her anonymous identity to avatar $B$ who obtains the iris template and chameleon public key from the blockchain according to the anonymous identity and throws a random challenge to $A$. Upon receiving the challenge, $A$ captures an iris sample from his/her manipulator to generate a signed iris sample as a response. If the iris sample matches the template and the chameleon collision signature matches the public key,  $B$ regards $A$ as legitimate.

\subsection{The Need Of Blockchain}
The metaverse is an open and long-lasting virtual ecosystem. The user's data must be carefully kept even if some platform operators withdraw. The blockchain has the advantages of public verification and prolonged storage. However, storing the first impression on the blockchain may reveal the users' privacy since the first impression contains social information \cite{Falchuk2018Privacy}. Therefore, the public verification and privacy preservation of the first impression must be resolved to support the intensive interactions in the metaverse \cite{Patwe2023User, Lin2022ICBCT}.

\subsection{Chameleon Collision Signature}
Because the traditional signature algorithms \cite{Thyagarajan2020Signatures, Boneh2004BLS} fail to represent the inner connection between two signature messages, it is difficult to verify the consistency between the avatar's virtual identity $VID$ and its physical identity $PID$. Chameleon signature \cite{Mohassel2010One} is a one-to-many signature mechanism, which signs multiple plaintexts using a signature hash. Based on this feature, the chameleon signature associates the avatar's $VID$ and $PID$ with the chameleon hash $h$. A new collision related to the chameleon hash \cite{Khalili2019Efficient} is forged by the chameleon private key, meaning that the new collision can be treated as a signature related to the old collision.

\begin{itemize}

\item $Setup(\mathcal{K})\rightarrow Parms$. The input of this probabilistic algorithm is a security parameter $\mathcal{K}$ and the output is the system parameter $Parms$.

\item $KeyGen(Parms)\rightarrow (pk,sk) $. The key generation algorithm takes the system parameter $Parms$ as the input and outputs the public-private key pair $(pk, sk)$.

\item $Hash(pk,M) \rightarrow (h,R)$. The hash algorithm takes $pk$ and a message $M$ as input and outputs the chameleon hash value $h$ and the check parameter $R$ of $M$.

\item $Check(pk,h,M,R)\rightarrow b\in\{0,1\}$. The compatibility detection algorithm takes as input the chameleon triplet$(h,M,R)$ and $pk$. It outputs a decision $b\in\{0,1\}$ indicating whether the $( pk, h, M, R )$ is compatible or not. 

\item $Sign(sk,h,M,R,M^\prime)\rightarrow R^\prime$. To sign a message $M^\prime$, the algorithm takes as input $sk$ and $(h,M,R)$. It outputs the check parameter $R^\prime$ of $M^\prime$, where the pairs $(M,R)$ and $(M^\prime, R^\prime)$ are called a colliding signature with respect to the hash value $h$.

\item $Verify(pk,h,M,R,M^\prime,R^\prime) \rightarrow b $.  To verify the colliding signatures $(M,R)$ and $(M^\prime, R^\prime)$ with respect to $h$, the algorithm detects the compatibility of $(h,M,R)$ and $(h,M^\prime,R^\prime)$ using the $Check$ algorithm. It outputs a decision $b\in\{0,1\}$ indicating whether the pairs $(M,R)$ and $(M^\prime, R^\prime)$ form a valid signature or not.
\end{itemize}

\subsection{Traditional Signcryption}
Signcryption  \cite{Zheng1997Signcryption}  is a cryptographic primitive and a typical signature-then-encryption technique guaranteeing the data conﬁdentiality and unforgeability. The traditional signcryption mechanism consists of the following four steps:

\begin{itemize}

\item $Setup(\mathcal{K})\rightarrow Parms$. The input of this probabilistic algorithm is a security parameter $\mathcal{K}$ and the output is the system parameter $Parms$.

\item $KeyGen(Parms)\rightarrow (pk,sk) $. The key generation algorithm takes the system parameter $Parms$ as input and outputs the public-private key pair $(pk, sk)$.

\item $SC(sk_A,M, pk_B) \rightarrow CT_{AB}$. The sender $A$ runs the signcryption algorithm by inputting his private key $sk_A$, a plaintext $M$, and the receiver's public key $pk_B$ to generate a ciphertext $CT_{AB}$ containing the signature parameter.

\item $DSC(pk_A,CT_{AB},sk_B)\rightarrow M \; \text{or} \perp$. The receiver $B$ executes the de-signcryption algorithm by inputting the sender's public key $pk_A$, the ciphertext $CT_{AB}$, and the receiver's private key $sk_B$ to recover the plaintext $M$. If an error occurs in retrieving $M$ from $CT_{AB}$,  the algorithm outputs $\perp$ to represent failure.

\end{itemize}

\section{Chameleon Signcryption} \label{sec:Signcryption}

In this section, we propose a signcryption mechanism with public verifiability based on the chameleon collision signature, called chameleon signcryption, to defend against the replacing attacks on first impressions.

\subsection{Some Definitions Of Chameleon Signcryption}
The proposed signcryption mechanism generates the signature associated with plaintext and ciphertext based on the chameleon collision to achieve public verifiability.  Inspired by Yang's chameleon collision signature \cite{Yang2023Trace} and Li's efficient signcryption mechanism \cite{Li2007EffiSC}, the signcryption mechanism is defined as follows.

\begin{itemize}

\item $Setup(\mathcal{K})\rightarrow Parms$.  The input of this probabilistic algorithm is a security parameter $\mathcal{K}$ and the output is the system parameter $Parms$.

\item $KeyGen(Parms)\rightarrow (pk,sk) $. The key generation algorithm takes the system parameter $Parms$ as the input and outputs the public-private key pair $(pk, sk)$.

\item $Hash(pk,M) \rightarrow (h,R)$. The chameleon hash algorithm takes as input the public key $pk$ and the message $M$. It outputs the chameleon hash value $h$ and the check parameter $R$ of $M$.

\item $Check(pk,h, M,R)\rightarrow b\in \{0, 1\}$. The compatibility detection algorithm takes as input $pk$ and the chameleon triplet $(h, M, R)$. It outputs a decision $ b \in \{0, 1\}$ indicating whether the $(h, M, R)$  is compatible or not.

\item $SC(sk_A,h$$_A$, $M^\prime,pk_B)\rightarrow CT_{AB}$. To generate a signcryption of $M^\prime$, the algorithm takes as input the sender's private key $ sk_A$,  the chameleon hash value $h$$_A$, and the receiver's public key $pk_B$. It outputs a ciphertext $CT_{AB}$ containing the signed message of $M^\prime$.

\item $VC(pk_A,CT_{AB}, h$$_A$, $\tilde{M},\tilde{R}, pk_B) \rightarrow b \in \{0,1\}$. The verification algorithm takes as input the sender's public key  $pk_A$, the ciphertext $CT_{AB}$, the chameleon triplet $(h$$_A$, $ \tilde{M},\tilde{R})$ and the receiver's $pk_B$. It outputs a decision $ b \in \{0, 1\}$ indicating whether the identity $(pk_A,pk_B)$ matches the ciphertext $CT_{AB}$ or not.

\item $DSC(pk_A,CT_{AB}, h$$_A$, $sk_B) \rightarrow (M^\prime,R^\prime) \; \text{or} \perp $.  The receiver executes this algorithm by inputting the sender's public key $pk_A$, the ciphertext $CT_{AB}$, the chameleon hash $h$$_A$, and the receiver's private key $sk_B$ to retrieve $(M^\prime,R^\prime)$ from $CT_{AB}$. If an error occurs, the algorithm outputs $\perp$.

\end{itemize}

\subsection{The Signcryption Process} \label{sec:CHSC}
The proposed signcryption $CH$-$SC=(Setup$, $KeyGen$, $Hash$, $Check$, $SC$, $VC$, $DSC)$ is constructed as follows:

\begin{itemize}

\item $Setup(\mathcal{K})\rightarrow Parms$. Let $\mathcal{K}$ be a security parameter in the signcryption system. The notations $\mathbb{G}$ and $\mathbb{G}_T$ are multiplicative cyclic groups of prime order $q\geq 2^\mathcal{K} $, where $g$ is a generator of $\mathbb{G}$ and the bit length of $g$ is $l$. The pairing  $e:\mathbb{G}\times\mathbb{G}\rightarrow \mathbb{G}_T$ is an eﬃciently computable bilinear map and $\mathbb{Z}_q$ is a finite field of order $q$, which satisfies $e(g^a,g^b)=e(g,g)^{ab}$  for any $a,b\in \mathbb{Z}_q$. The system selects three global anti-collision hash functions $H_1:\{0,1\}^*\rightarrow\mathbb{G}$, $H_2:\mathbb{G} \times \mathbb{G} \times \mathbb{G} \rightarrow \{0,1\}^{n+l}$, and $H_3:\mathbb{G}  \times \{0,1\}^{n+l} \times \mathbb{G} \rightarrow \{0,1\}^n$, where $H_1$ mapping bit strings of arbitrary length to an elements in $\mathbb{G}$, $H_2$ and $H_3$ are similar to $H_1$ except that its input and output elements are different. Finally, the system publishes the parameter $Parms=\{\mathbb{G},\mathbb{G}_T,g_,q,e,H_1, H_2, H_3\}$.

\item $KeyGen(Parms)\rightarrow (pk,sk)$. The key generation algorithm takes the system parameter $Parms$ as the input. The algorithm picks a random value $x\stackrel{R}{\leftarrow}  \mathbb{Z}_q$ as the private key $sk$ and calculates $y=g^x\in \mathbb{G}$ as the public key $pk$, where the symbol ``$\stackrel{R}{\leftarrow}$'' means to randomly select an element from a set. The algorithm outputs the public-private key pair $(pk,sk)$ as	
$$sk = x, \; pk = y. $$

\item $Hash(pk,M) \rightarrow (h,R)$. The algorithm takes as input the public key $pk=y$ and the message $M\in \{0,1\}^n$. It outputs the chameleon hash value $h$ and the corresponding check parameter $R$ of $M$ as
\begin{equation}\nonumber \label{equ:Hash}
\begin{split}
     m=H_1(M),\; r\stackrel{R}{\leftarrow} \mathbb{Z}_q,\\
     h=m \cdot y^r,  \; 	R=g^r.
\end{split}
\end{equation}
\item $Check(pk,h,M,R)\rightarrow b \in \{0,1\}$. The algorithm takes as input the public key $pk=y$ and the chameleon triple $(h,M,R)$.  It outputs $b\in \{0,1\}$ as
\begin{align}
        m=H_1(M),\nonumber \\  
      e(h/m,g) \stackrel{?}{=} e(R,y).  \tag{4.1} \label{eq:Check}  
\end{align} 
Among them, the symbol ``$\stackrel{?}{=}$'' indicates whether the equation is hold or not. If the equation \eqref{eq:Check} holds, the algorithm outputs $b=1$; otherwise $b=0$. 

\item $SC(sk_A,h$$_A$, $M^\prime,pk_B)\rightarrow CT_{AB}$. To generate a signcryption of $M^\prime \in \{0,1\}^n$, the algorithm takes as input the sender's private key $ sk_A=x_A$, the chameleon hash value $h$$_A$, and the receiver's public key $pk_B=y_B$. It outputs the ciphertext $CT_{AB}$ as
\begin{equation} \nonumber
\begin{split}
     k \stackrel{R}{\leftarrow} \mathbb{Z}_q, \; K=g^k, \\
    m^\prime=H_1(M^\prime),\; R^\prime=(h{_A}/m^\prime)^{(1/x_A)},\\
    Z=(M^\prime||R^\prime) \oplus H_2(K,y_B,y_B^k),\\    M^{\prime\prime}=H_3(K,Z,y_B),\;m^{\prime\prime}=H_1(M^{\prime\prime}),\\
    R^{\prime\prime}=(h{_A}/m^{\prime\prime})^{(1/x_A)}, \;
    CT_{AB}=(K,Z,R^{\prime\prime}).
\end{split}
\end{equation}

In the above formula, ``$||$'' means concatenating bit strings.

\item $VC(pk_A,CT_{AB},h${$_A$}, $\tilde{M},\tilde{R}, pk_B) \rightarrow b \in \{0,1\}$. The verification algorithm of ciphertext takes as input the sender's  $pk_A=y_A$, the chameleon triplet $(h${$_A$}, $\tilde{M},\tilde{R})$, the ciphertext $CT_{AB}$, and the receiver's $pk_B=y_B$. It outputs a decision $ b \in \{0, 1\}$ as
\begin{align}
        (K,Z,R^{\prime\prime}) \leftarrow CT_{AB}, \notag\\
       Check(pk_A,h{{_A}},\tilde{M},\tilde{R}) \stackrel{?}{=} 1, \tag{4.2} \label{eq:VC1} \\ 
       Check(pk_A,h{{_A}},H_3(K,Z,y_B),R^{\prime\prime})  \stackrel{?}{=} 1. \tag{4.3} \label{eq:VC2} 
\end{align}
If both the equations \eqref{eq:VC1} and \eqref{eq:VC2} hold, the algorithm outputs $b=1$; otherwise $b=0$.

\item $DSC(pk_A,CT_{AB},h${$_A$}, $sk_B) \rightarrow (M^\prime, R^\prime) \; \text{or}  \perp $.  The algorithm retrieves $(M^\prime,R^\prime)$ from ciphertext $CT_{AB}$ as
\begin{align}
    (K,Z,R^{\prime\prime}) \leftarrow CT, \notag \\
    Check(pk_A,h{{_A}},H_3(K,Z,y_B),R^{\prime\prime}) \stackrel{?}{=} 1,\tag{4.4} \label{eq:DSC1} \\
    (M^\prime||R^\prime)=Z \oplus H_2(K,y_B,K^{x_{_B}}),\notag \\
   Check(pk_A,h{{_A}},M^\prime,R^{\prime}) \stackrel{?}{=} 1. \tag{4.5} \label{eq:DSC2}
\end{align}
 If the equations \eqref{eq:DSC1} and \eqref{eq:DSC2} hold, the algorithm outputs $(M^\prime, R^\prime)$; otherwise it outputs $\perp$ to represent failure. 
\end{itemize}

\section{Anti-disguise Authentication Framework} \label{sec:Framework}
\begin{figure}[htbp]
\begin{center}
    \includegraphics[width=0.47\textwidth]{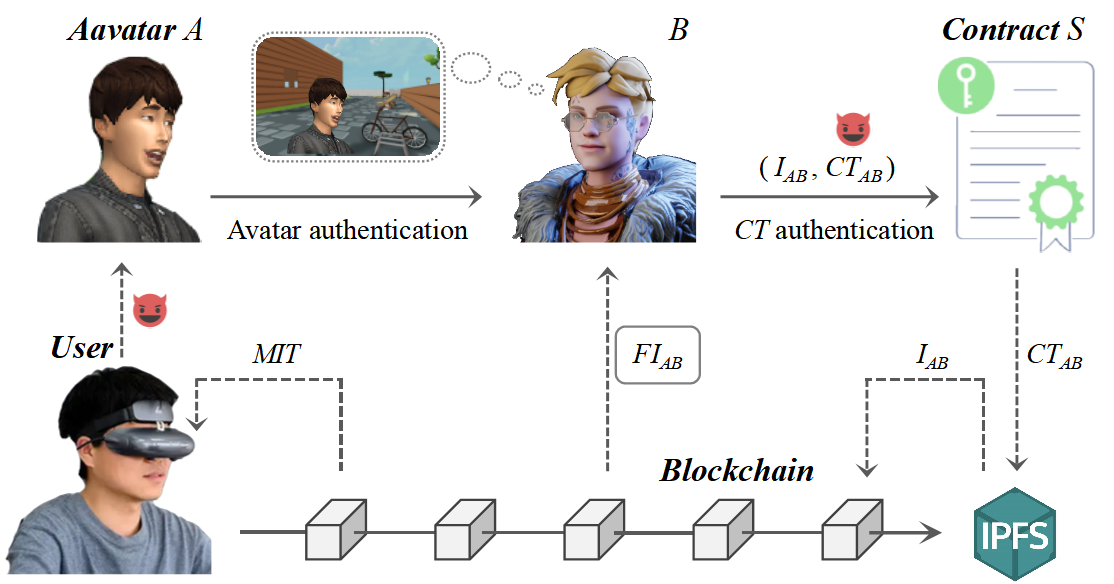}
    \caption{\small{In the system framework, 
 we only show the process that $B$ writes $(I_{AB}, CT_{AB})$ to form $FI_{AB}$. In fact, $A$ writes $(I_{BA}, CT_{BA})$ in the same way as $B$ to form $FI_{BA}$.}}
    \label{fig:FrameWork}
 \end{center}
\end{figure}

In this section, we construct an authentication framework to defend against the disguise attack based on first impression while protecting the first impression from being replaced and forged. During the authentication process between avatar $A$ and avatar $B$, the verifier $B$ extracts the first impression $FI_{AB}$ from the blockchain to perceive $A$'s core identity, which prevents an adversary $C$ from mounting the disguise attack. If the first impression $FI_{AB}$ is empty, meaning that this is the first meeting between $A$ and $B$, they create the first-impression identity and store it in the blockchain, respectively. The system framework is shown in Fig.\ref{fig:FrameWork}.

\subsection{System Framework}

\begin{itemize}

\item $\boldsymbol{User}$ is the physical player manipulating an avatar. Before entering the metaverse, he/she needs to register his/her identity in an identity provider (IDP) to obtain a metaverse identity token (MIT), which allows the user to create an appearance as his own avatar.

\item $\boldsymbol{Avatar}$ is the virtual appearance of a physical user in the metaverse that is expected to only represent the real-world manipulator (i.e., one-to-one mapping).

\item  $\boldsymbol{Contract}$ provides interfaces for entities in the metaverse to authenticate the avatar and the ciphertext (CT) of first impression (FI).

\item $\boldsymbol{Blockchain}$ stores public parameters related to users' identities. To reduce the storage cost, we introduce IPFS to store files such as $MIT$ and $CT$ and blockchain to store the extraction index $I$, which forms the first impression $FI=(I,CT)$.

\end{itemize}

\subsection{Security Threats}
To mislead $B$ to treat an attacker as the friend $A$, the adversary $C$ executes the \textit{replacing} and \textit{forging} attacks on the first impression $FI_{AB}$ to achieve the \textit{disguising} process. Moreover, considering the privacy issue, the adversary may intend to infer the user's sensitive and identifiable information based on the data stored on blockchain and IPFS.

\begin{itemize}

\item \textbf{Replacing Attack} refers to that the adversary $C$ replaces the ciphertext $CT_{CB}$ of $FI_{CB}$ with $CT_{AB}$ in the storage process, which misleads $B$ to either recall the first impression of $A$ by $C$'s identity, or fail to parse the first impression.

\item \textbf{Forging Attack} refers to that the adversary $C$ generates a new signcryption $CT_ {CB}^\prime$ and stores it in blockchain to form the first impression $FI_{CB}=(I_{CB},CT_{CB}^\prime)$, which means that $C$ creates an illegal or non-negotiable first impression.

\item \textbf{Disguising Attack} refers to that the adversary $C$ generates the same avatar as $A$ to mislead $B$, where the $FI_{CB}$ may have been replaced or forged.

\item  \textbf{Privacy-inference Attack} refers to the adversary collecting the user's personal data such as $CT$ in $FI$ and $MIT$ from IPFS and blockchain, revealing the first metaverse-meeting scene (i.e., the first impression) or the unique identity.

\end{itemize}

\section{Anti-disguise Authentication Protocol} \label{sec:Protocol}
The first impression can be used to realize the anti-disguise authentication for avatars. However, the first impression faces the threat of replacing and forging attacks. In this section, we first design a user's identity model with the first impression in the metaverse. Then, we design two authentication protocols, where the avatar authentication protocol based on the first impression is to defend against the disguise attack, and the ciphertext authentication protocol based on the chameleon signcryption is to avoid the replacing and forging attack on the first impression. 
The related symbols are presented in TABLE \ref{tab:Notatins}.

\begin{table}[htbp]
  \begin{center}
  \caption{Avatar's Identity Parameters and Corresponding Meanings}
  \label{tab:Notatins}
  \begin{tabular}{ccc}
        \hline
        \specialrule{0em}{1pt}{1pt}
         Symbol & Description   \\
        \specialrule{0em}{1pt}{1pt}
        \hline     
        \specialrule{0em}{1pt}{1pt}
          $ID$ & The user's identification number $ID=(Rid,Mid)$ \\
          $Rid$ & The unique identity in the real world  \\ 
         $Mid$ & The unique identity in the metaverse \\ 
         $MIT$ & The metaverse identity token $MIT=(SN,$ $ Hid$, $pk, T)$ \\ 
          $Hid$ & The anonymous ID corresponding to $Mid$  \\
          $SN$ & The serial number of a $MIT$ \\
          $ pk $ & The user's chameleon public key \\
          $T $ & The user's  iris template \\
        $ Avatar $ & The digital appearance $Avatar=(Hid$, $h$, $VID, PID)$ \\
         $h$ &   The avatar's chameleon hash \\
         $ VID $ &  The avatar's virtual identity \\
         $M_a$ & The visible identity of an avatar\\
         $R_a$ & The check parameter of $M_a$\\
        $ PID $ & The avatar's physical identity \\
        $M_a^\prime$ & An iris sample of the avatar's manipulator \\
        $R_a^\prime$ & The check parameter of $M_a^\prime$\\
        $Img_A$ & The image containing avatar $A$'s appearance \\
        $FI_{AB}$ & The first impression $A$ make for $B$  \\
        $I_{AB}$ & The extraction index for $CT_{AB}$ \\
        $CT_{AB}$ & The ciphertext about a first metaverse-meeting scene \\
         \specialrule{0em}{1pt}{1pt}
        \hline
  \end{tabular}
  \end{center}
\end{table}

\subsection{The User's Identity Model}
\begin{figure}[htbp]
\begin{center}
    \includegraphics[width=0.43\textwidth]{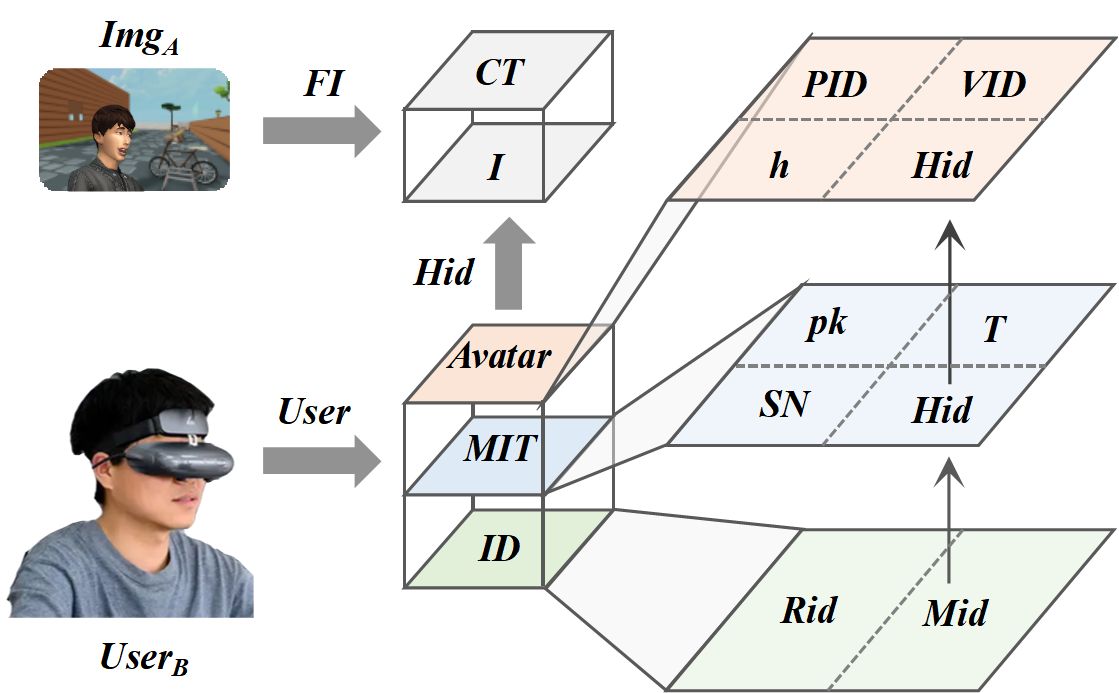}
    \caption{\small{The user's identity model in the metaverse.}}
    \label{fig:IdenMod}
 \end{center}
\end{figure}

The user's identity model $User=\{ID,MIT,Avatar\}$ (as shown in Fig.\ref{fig:IdenMod}) is a security enhancement measure over the avatar's identity model \cite{Yang2023Trace}. 

The user's identification number $ID=(Rid,Mid)$ includes the unique identity $Rid$ in the real world and the corresponding unique identity $Mid$ in the metaverse, where the $Mid$ is treated as the core identity of an avatar as shown in TABLE \ref{tab:Identities}. {Considering that the metaverse users come from different countries and districts, we put the country and district codes in the core identity $Mid$. To avoid conflicts, we introduce time and personal serial numbers (PSN) into $Mid$. Therefore, the $Mid$ is constructed by 25 decimal digits according to TABLE  \ref{tab:FieldsMid}. For example, an entity registered in Beijing, China on June 1, 2024, can be expressed as $Mid_A =156\; 110105\; 20240601\; 301107$ in which the country code 156 in ISO 3166 represents China and the district code 110105 represents Beijing.}

\begin{table}[htbp]

      \centering
      \caption{User's Identities in Metaverse}
      \label{tab:Identities}
      \begin{threeparttable}
      \begin{tabular}{p{60 pt}<{\centering}p{40 pt}<{\centering}p{40 pt}<{\centering}p{60 pt}<{\centering}}
          \hline
        \specialrule{0em}{1pt}{1pt}
        Identities & Symbol & Belong to  & Implying in  \\
        \specialrule{0em}{1pt}{1pt}
           \hline
          \specialrule{0em}{1pt}{1pt}
          Core Identity  &  $Mid$, $Hid$ &  $User$ & $ID$ and $FI$ \\
          Virtual Identity & $VID$ &  $Avatar$  & -- \\
          Visible Identity  &  $M_a$ & $Avatar$    &  $VID$ \\
          Physical Identity &  $PID$  & $Avatar$   & -- \\
          Owner's Identity &   $(K,Z),pk$  & $FI$     & $CT$ and $MIT$ \\
          Writer's Identity &  $Hid,pk$  & $FI$ &  $I$ and  $MIT$   \\
      \specialrule{0em}{1pt}{1pt}
        \hline
      \end{tabular}
\end{threeparttable}
\end{table}

\begin{table}[htbp]
      \footnotesize
      \centering
      \caption{Field and the corresponding length in $Mid$}       \label{tab:FieldsMid}
      \begin{threeparttable}
      \begin{tabular}{p{70 pt}<{\centering}p{30 pt}<{\centering}p{30 pt}<{\centering}p{30 pt}<{\centering}p{30 pt}<{\centering}}
          \hline
        \specialrule{0em}{1pt}{1pt}
       Field of $Mid$ & Country  & District & Date & PSN \\
        \specialrule{0em}{1pt}{1pt}
           \hline
          \specialrule{0em}{1pt}{1pt}
         Length of numbers  &  3  & 6 & 8 & 6 \\
        \specialrule{0em}{1pt}{1pt}
        \hline
      \end{tabular}
\end{threeparttable}
\end{table}

{The metaverse identity token $MIT$ is a bridge to connect the real-world user and his/her avatar in the metaverse. Before entering the metaverse, the user generates the avatar based on his/her anonymous identity token $MIT=(SN,$ $ Hid$, $pk, T)$, where $SN$ is the serial number of $MIT$,  the hashed string $Hid$ is generated using the hash algorithm (e.g., SHA256) to realize anonymity,  $pk$ is the user's public key, and $T$ is the biometric template extracted from the user's iris.}

The user's digital appearance $Avatar=(Hid$, $h$, $VID$, $PID)$ consists of the avatar's chameleon hash $h$ extracted from $(h,R_a) \leftarrow Hash(pk,M_a)$, the public virtual identity $VID=(M_a ,R_a)$ , and the traceable physical identity $PID=(M_a^\prime,R_a^\prime)$.  In $VID$, the variable $M_a$ is the visible identity of the avatar which is shown to others in the metaverse, $R_a$ is the check parameter of $M_a$. In $PID$, $M_a^\prime$ is an iris sample  \cite{John2020Eye} extracted from the avatar's manipulator, $R_a^\prime$ is the check parameter of $M_a^\prime$. 

To prevent the adversary from mounting the disguise attack, we create the ``first impression'' of an avatar based on its $Hid$. For example, the first impression $A$ make for $B$ is created as $FI_{AB}=(I_{AB},CT_{AB})=($ $Hid_A| |Hid_B$, $CT_{AB})$, where $I_{AB}$ is the extraction index and $CT_{AB}$ is the ciphertext generated by the first metaverse-meeting scence $Img_A$.

\begin{figure*}[htbp]
    \centering
     \subfigure[Avatar authentication protocol]{
     		\label{fig:AvatarAuthProt}    \includegraphics[width=8.5cm]{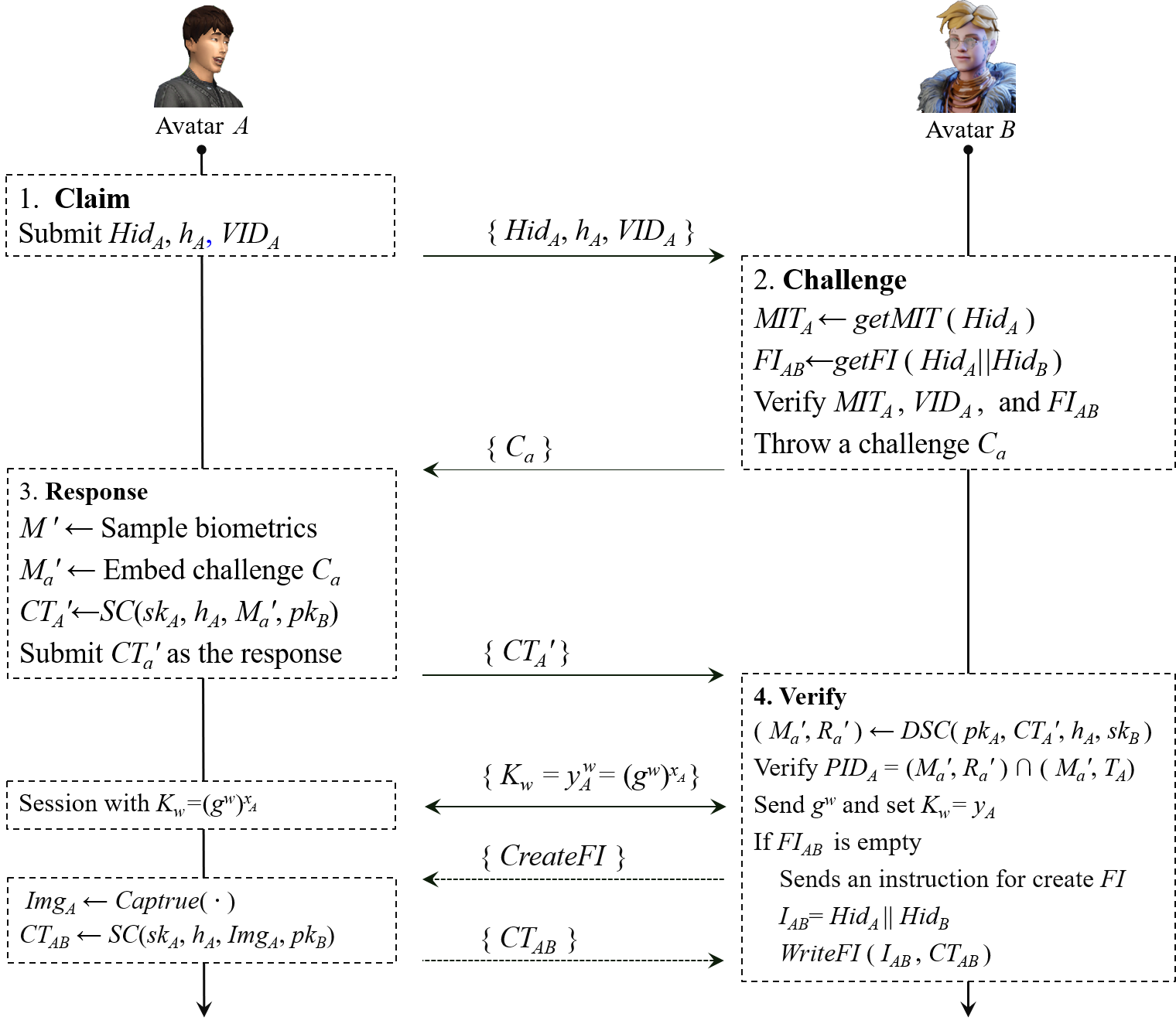}    }    \quad
    \subfigure[Ciphertext authentication protocol]{
    		\label{fig:CryptAuthProt}    \includegraphics[width=8.5cm]{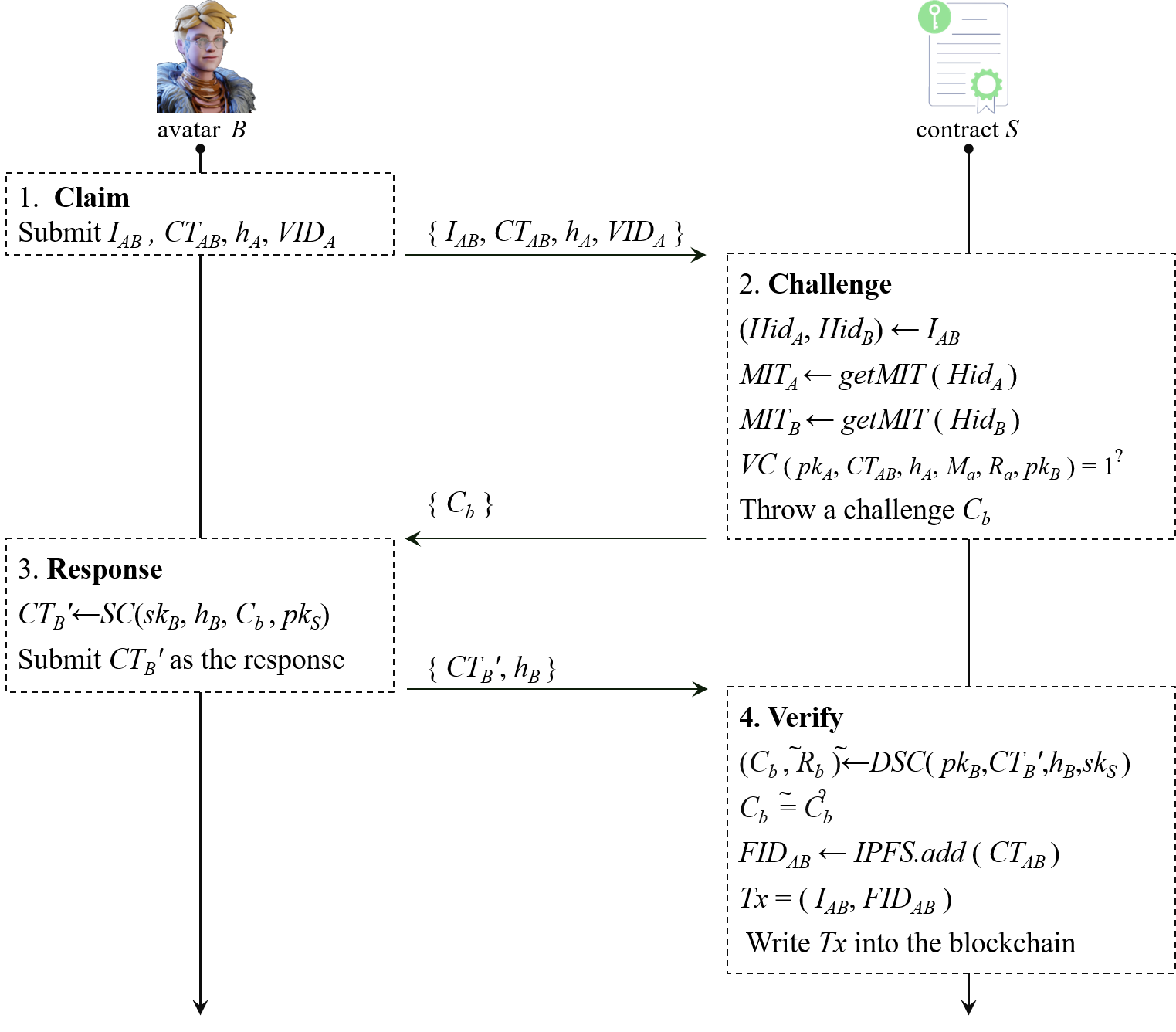}    }    \quad
    \caption{The avatar authentication protocol and the ciphertext authentication protocol.}
    \vspace{-1.1em}
\end{figure*}

\subsection{Avatar Authentication Protocol} \label{sec:AvtAuthPro}

When two avatars meet at a place in the metaverse, they verify each other's identities based on first impressions to defend against the disguise attack. For simplicity, we utilize avatar $B$ as the verifier to verify avatar $A$'s identity. The formal description of the avatar authentication protocol is shown in Fig. \ref{fig:AvatarAuthProt}. The protocol is described as follows: i)  $A$ acting as a prover provides $B$ with {$Hid_A, h_A,$} and $VID_A$  to claim that the identity of $A$ is valid; ii) $B$ acting as a verifier checks $A$'s $MIT_A$, $VID_A$, and $FI_{AB}$, and then throws a random challenge $C_a$ to $A$, which confirms whether $A$'s physical identity is consistent with his/her virtual identity or not; iii) $A$'s manipulator provides the ciphertext $CT_A^\prime$ containing his/her physical identity as a response to $B$; iv) $B$ decrypts $CT_A^\prime$ and checks the parameters to determine whether the virtual and physical identities are consistent or not. If $FI_{AB}$ is empty, meaning that the two avatars meet for the first time, then $B$ sends an instruction to $A$ for creating the first impression. Details are as follows:

\subsubsection{\textbf{Claim}}
First, $A$ provides $B$ with {$Hid_A, h_A$} and $VID_A=(M_a,R_a)$ to initiate an identity claim.

\subsubsection{\textbf{Challenge}}
Next, $B$ verifies $A$'s virtual identity by the following three steps: (i) $B$ obtains $MIT_A$ from the blockchain according to {$Hid_A$}; (ii) $B$ sets {$I_{AB}=Hid_A||Hid_B$} to get $FI_{AB} =(I_{AB}, CT_{AB})$ from blockchain and IPFS; (iii) $B$ verifies IDP's signature on $MIT_A$ to ensure the validity of $MIT_A$, verifies $VID_A=(M_a,R_a)$ by $Check(pk_A, h_A, M_a, R_a)$ to ensure the validity of $A$'s visible identity, and decrypts the ciphertext $CT_{AB}$ and reviews the first metaverse-meeting scene $Img_A \leftarrow DSC(pk_A,h_A,CT_{AB},sk_B)$ to defense against disguise attack in an unconscious way. If the above three steps are passed, $B$ throws a random challenge $C_a$ to $A$.

\subsubsection{\textbf{Response}}
$A$ submits its physical identity by the following four steps: (i) $A$’s manipulator generates the iris feature $M^\prime$; (ii) $A$ embeds $C_a$ into $M^\prime$ to form a marked iris feature $M_a^\prime$; (iii) $A$ generates a ciphertext $CT_A^\prime \leftarrow SC(sk_A$, $h_A, M_a^\prime, pk_B)$ and sends it to $B$ as the response to $C_a$.

\subsubsection{\textbf{Verify}}
Next, $B$ decrypts $CT_A^\prime$ to construct $PID_A=(M_a^\prime,R_a^\prime) \leftarrow DSC(pk_A,h_A,CT_A^\prime,sk_B)$  and verifies $PID_A$ through the following three steps : (i) $B$ extracts the watermark $C_a^\prime$ from $M_a^\prime$ and detects $C_a^\prime \stackrel{?}{=} C_a$ to determine the freshness of $M_a^\prime$;  (ii) $B$ verifies the match between collisions  $PID_A = (M_a^\prime,R_a^\prime)$ and {$VID_A=(M_a,R_a)$} by $Check(pk_A, h_A, M_a^\prime , R_a^\prime)$; (iii) $B$ verifies the match between the iris feature $M_a^\prime$ and {$T_A$} in $MIT_A$ to determine the validity of $A$'s physical identity. If the above three steps are passed, $B$ randomly selects $w \stackrel{R}{\leftarrow} \mathbb{Z}_q$ and sends $g^w$ to $A$ by which builds a session key $K_w=y_A^w=(g^w)^{x_{_A}}$.

If the $FI_{AB}$ extracted in the challenge phase is empty, $B$ sends $A$ an instruction to ask $A$ to create the first impression $FI_{AB}=(I_{AB}, CT_{AB})=(${$Hid_A||Hid_B$}, $CT_{AB})$ which be written into the blockchain and IPFS. Among them, the ciphertext $CT_{AB} \leftarrow SC(sk_A,h_A,Img_A,pk_B)$ is generated by a snapshot image $Img_A$ containing $A$'s facial appearance and the metaverse-meeting scene.

It is worth noting that both $A$ and $B$ act as verifiers to authenticate each other's identities when they meet. Therefore, $A$ also authenticates $B$ based on the protocol as Fig.\ref{fig:AvatarAuthProt} and sends $B$ an instruction to ask $B$ to create a first impression $FI_{BA}$ when they first meet.

\subsection{Ciphertext Authentication Protocol} \label{sec:CipAuthProt}
Based on the real-world social case, the first impression in the metaverse can only be written by the owner himself/herself when storing the first impressions. To avoid the replacing and the forging attacks against the first impression, we propose a ciphertext authentication protocol based on the public verifiability of chameleon signcryption, which guarantees that the owner's identity implied in the ciphertext is consistent with the writer's identity implying index. The ciphertext authentication protocol is illustrated in Fig.\ref{fig:CryptAuthProt}. The protocol is described as follows: i) $B$ acting as a prover sends $\{I_{AB}, CT_{AB}$, {$h_A,VID_A$}$\}$ to the smart contract $S$ to claim that he/she is the legal writer of $CT_{AB}$; ii) $S$ acting as a verifier checks the owner's identity implying in $CT_{AB}$ and throws a random challenge $C_b$ to $B$; iii) $B$ returns $S$ a response $\{CT_B^\prime$, {$h_B$}$\}$ to prove the validity of the writer's identity implied in $I_{AB}$; iv) $S$ decrypts $CT_B^\prime$ and checks the corresponding parameters to determine whether the writer's identity is consistent with the owner's identity or not. If all checks are validated, $S$ stores $CT_{AB}$ in IPFS and writes $I_{AB}$ in the blockchain to construct the first impression $FI_{AB}= (I_{AB},CT_{AB})$. The specific authentication process is described as follows.

\subsubsection{\textbf{Claim}}
First, $B$ provides $S$ with $\{I_{AB}, CT_{AB}$, {$h_A,VID_A$}$\}$ to claim that he/she is the legal writer of $CT_{AB}$.

\subsubsection{\textbf{Challenge}}
$S$ checks the owner's identity through the following three steps: (i) $S$ parses  $I_{AB}=$ {$Hid_A||Hid_B$} to get {$MIT_A=(SN_A, Hid_A,pk_A,T_A)$} and {$MIT_B=(SN_B$, $Hid_B$, $pk_B,T_B)$}; (ii) $S$ utilizes $pk_B$ and $VC(pk_A$, $ CT_{AB}$, {$h_A$,  $M_a$, $R_a$}, $pk_B)$ to verify the owner's identity $B$ implying in ciphertext $CT_{AB}$. If the verification is passed, $S$ throws a random challenge $C_b$ to $B$ to guarantee that the writer's identity is consistent with the owner's identity.

\subsubsection{\textbf{Response}}
$B$ first generates a signcryption $CT_B^\prime \leftarrow SC(sk_B,h_B,C_b,pk_S)$ using its private key $sk_B$ and $S$'s public key $pk_S$, then submits $\{CT_B^\prime$, {$h_B$}$\}$ to $S$ as a response for $C_b$. Among them, the public key $pk_B$ corresponding to $sk_B$ indicates the writer's identity $B$ implying in index $I_{AB}$. 

\subsubsection{\textbf{Verify}}
$S$ extracts $(\tilde{C}_b,\tilde{R}_b) \leftarrow DSC(pk_B, h_B, CT_B^\prime, sk_S)$ based on $pk_B$ and verifies the validity of $(\tilde{C}_b,\tilde{R}_b) $ through $\tilde{C}_b \stackrel{?}{=} C_b$, which guarantees the consistency between the writer's identity implying in $I_{AB}$ and the owner's identity implying $CT_{AB}$. If the verification is passed, $S$ first adds $CT_{AB}$ to IPFS to obtain the file number $FID_{AB}$, then constructs a transaction $Tx=(Key,Value)=(I_{AB}$, $FID_{AB})$ and writes it in the blockchain. The blockchain stores the index $(I_{AB}, FID_{AB})$ while IPFS stores the file $(FID_{AB}, CT_{AB})$, which forms the first impression $FI_{AB}=(I_{AB},CT_{AB})$.

\section{Security Analysis} \label{sec:SecuAnaly}
The goal of this work is to defend against the disguise attack on the avatar's appearance in the metaverse based on the first impression. However, the ciphertext of the first impression suffers from the replacing and the forging attacks. The capability of the ciphertext authentication protocol against the replacing attack depends on the public verifiability and the unforgeability of the chameleon signcryption mechanism. Thus, we first analyze the security of the signcryption mechanism, then analyze the security of the authentication protocol.

\subsection{Signcryption Security}

The signcryption mechanism adds a ciphertext verification algorithm to achieve public verifiability. The designed signcryption needs to satisfy the ciphertext unforgeability and the public verifiability except for the traditional signcryption unforgeability and confidentiality \cite{Li2007EffiSC}. 

\subsubsection{SC-EUF-CMA}
The real-attack environment for forging a signcryption is that the adversary $\mathcal{A}$ has the user $\mathcal{B}$ to generate the signcryption about arbitrary plaintext $M$ to any receiver $Y_R$. The adversary's purpose is to forge a signcryption $CT^*$ of a new plaintext $M^*$ such that the signcryption can pass the decryption and verification. Thus, the ideal-attack environment is that $\mathcal{A}$ controls the signcryption oracle to output a forged signcryption $CT^*$ about $M^*$ after inquiring the oracle for a certain number of times. If  $CT^*$ is able to pass de-signcryption correctly, $\mathcal{A}$ successfully forges a signcryption; otherwise $\mathcal{A}$ fails. The formal description of the attack is shown as the game $Exp_{CH\text{-}SC,\mathcal{A}}^{SC-EUF}(\mathcal{K})$ given in the Appendix  A-A of the full paper \footnote{ https://github.com/zzyzhuiying/zzyzhuiying.github.io/blob/main/Full.pdf}. The signcryption unforgeability can be reduced to the divisible computation Diiﬁe-Hellman (DCDH) problem\cite{Bao2003Variations}. Here, we show that even if the adversary $\mathcal{A}$ controls the oracle, he/she doesn't have obvious advantages in forging signcryption $CT^*$ about $M^*$. The detailed proofs are given in the Appendix of \ref{sec:ProofScEuf}.

\textbf{Definition 1 (Signcryption Unforgeability).} We say that the chameleon signcryption satisfies unforgeability under adaptive chosen message attacks (SC-EUF-CMA), if there is no polytime adversary who wins the game $Exp_{CH\text{-}SC,\mathcal{A}}^{SC-EUF}(\mathcal{K})$ with a non-negligible advantage.

\textbf{Theorem 1}. If the DCDH assumption holds on $\mathbb{G}$, the chameleon signcryption satisfies SC-EUF-CMA.

\subsubsection{C-EUF-CMA}
The real-attack environment of forging ciphertext is similar to forging signcryption. The difference is that the forged ciphertext doesn't need to be associated with a plaintext. The formal description of the attack process is shown as the game $Exp_{CH\text{-}SC,\mathcal{A}}^{C-EUF}(\mathcal{K})$ given in the Appendix A-B of the full paper.  Here, we show that $\mathcal{A}$ doesn't have obvious advantages in forging ciphertext $CT^*$. The detailed proofs are given in the Appendix of \ref{sec:ProofCEuf}.

\textbf{Definition 2 (Ciphertext Unforgeability).} We say that the chameleon ciphertext satisfies unforgeability under adaptive chosen message attacks (C-EUF-CMA), if there is no polytime adversary who wins the game $Exp_{CH\text{-}SC,\mathcal{A}}^{C-EUF}(\mathcal{K})$ with non-negligible advantage.

\textbf{Theorem 2.} If the DCDH assumption holds on $\mathbb{G}$, the proposed signcryption mechanize satisfies C-EUF-CMA.

\subsubsection{SC-IND-CCA}
The real-attack environment of confidentiality is that the adversary $\mathcal{A}$ lets the user $U$ generate arbitrary ciphertext about $M$ to any receiver $Y_R$ or lets $U$ decrypt the ciphertext $CT$ receiving from any sender $Y_S$. The adversary's purpose is to obtain part of the plaintext information, such as the 1-bit plaintext, from a new ciphertext sent by $U$. The ideal-attack environment is that the adversary $\mathcal{A}$ controls the signcryption oracle $\mathcal{SC}$ and the designcryption oracle $\mathcal{DSC}$ to distinguish the source of ciphertext $CT^*$, which is generated from one of the known plaintexts $m_0$ and $m_1$. The formal description of the attack shows as the game $Exp_{CH\text{-}SC,\mathcal{A}}^{SC-CCA}(\mathcal{K})$  {given in the Appendix A-C of the full paper}.  The security of signcryption confidentiality can be reduced to the computational Diﬃe-Hellman (CDH) problem \cite{Bao2003Variations}. Here, we obtain that even if the adversary $\mathcal{A}$ controls the oracles, he/she doesn't have obvious advantages in distinguishing  $CT^*$.  The detailed proofs are given in the Appendix of \ref{sec:ProofCCA}.

\textbf{Definition 3 (Signcryption Confidentiality).} We say that the signcryption mechanism satisfies semantically secure under adaptive chosen ciphertext attack (SC-IND-CCA), if there is no polytime adversary who wins the game $Exp_{CH\text{-}SC,\mathcal{A}}^{SC-CCA}(\mathcal{K})$ with a non-negligible advantage.

\textbf{Theorem 3.} If the CDH assumption holds on $\mathbb{G}$, the proposed signcryption mechanism satisfies SC-IND-CCA.

\subsubsection{Public Verifiability}
Public verifiability means that the identities of the sender and the receiver's implied in a ciphertext can be verified without decryption. In the proposed signcryption mechanism, anyone is able to verify the identities $A$ and $B$ implying in $CT_{AB}$ without decryption based on $VC (pk_A , CT_{AB}, h_A,\tilde{ M}_A,\tilde{R}_A, pk_B )$. Considering the attack, the adversary may construct public key and chameleon parameters to mislead the verifier, resulting in incorrect recognitions of the sender's or receiver's identity. The public verifiability depends on the CDH assumption and the collision resistance of $H_3$.  The detailed proofs are given in the Appendix of \ref{sec:ProofPubVeri}.

\textbf{Theorem 4.} If the CDH assumption holds on $\mathbb{G}$ and $H_3$ satisfies collision resistance, the proposed signcryption mechanism satisfies public verifiability.

\subsection{Protocol Security}

The goal of the avatar authentication protocol is to defend against the disguise attack, while that of the ciphertext authentication protocol is to defend against the replacing and the forging attack on the first impression. In the following, the security of these two protocols is deeply analyzed.

\subsubsection{Defending against the replacing attack}
The replacing attack is constructed as an adversary $\mathcal{C}$ utilizes $A$'s ciphertext in the first impression to replace $\mathcal{C}$'s ciphertext in the process of written first impression, leading to the verifier parsing out the first impression of $A$ using $\mathcal{C}$'s identity. The defense against the ciphertext replacing attack depends on the public verifiability of the chameleon signcryption. 

If $\mathcal{C}$ attempts to mount a replacing attack when $B$ writes the first impression $FI_{CB}=(I_{CB},CT_{CB})$ in the blockchain, then $\mathcal{C}$ must replace the $CT_{CB}$ in $FI_{CB}$ with $CT_{AB}$ to form $FI_{CB}= (I_{CB},CT_{AB})$. In the challenge phase of the ciphertext authentication protocol, since the signcryption satisfies public verifiability, the contract $S$ extracts the public keys $pk_C$ and $pk_B$ as identities $\mathcal{C}$  and $B$ according to the index $I_{CB}=$ {$Hid_C||Hid_B$} to verify the identities implied in $CT_{AB}$ according to $VC(pk_C$, {$CT_{AB},h_C,M_C,R_C$,} $pk_B) \neq 1$, indicating that $\mathcal{C}$ and $B$ do not match $CT_{AB}$. Thus, the first impression $FI_{CB}=(I_{CB},CT_{AB})$ replaced by $\mathcal{C}$ cannot pass the ciphertext authentication. 

\subsubsection{Defending against the forging attack}
The forging attack on the first impression is constructed as an adversary $\mathcal{C}$  generates a ciphertext and writes it into blockchain to form a first impression of $\mathcal{C}$. The defense against the forging attack depends on the SC-EUF-CMA of the chameleon signcryption.

If $\mathcal{C}$ wants to mount the forging attack using $A$’s first metaverse-meeting scene, then $\mathcal{C}$ must generate the ciphertext $CT_{CB}$ and the index $I_{CB}$ to form $FI_{CB}=(I_{CB},CT_{CB})$ such that $FI_{CB}$ is able to pass the owner verification and the writer verification. In the challenge phase of the ciphertext authentication protocol, the contract $S$ checks $CT_{CB}$ to verify the owner's identity. Since $CT_{CB}$ is constructed by $\mathcal{C}$'s private key and $B$'s public key, the owner's identity $B$ implying in $CT_{CB}$ satisfies $VC(pk_C,CT_{CB},h_C$, {$M_C$}, $R_C,pk_B)=1$, where the $pk_B$ is treated as the owner's identity. In the verify phase, $S$ decrypts  $CT_C^\prime$ and checks $\tilde{C}_b$ to verify the writer's identity. Since $\mathcal{C}$ does not have the private key of $B$ while the proposed chameleon signcryption satisfies SC-EUF-CMA, the $CT_C^\prime$ generated by $\mathcal{C}$ fails to cannot pass the verification. That is, $\tilde{C}_b \neq C_b$, where $\tilde{C}_b$ is extracted from $ (\tilde{C}_b,\tilde{R}_b) \leftarrow DSC(pk_B,CT_C^\prime, h_B,sk_S)$, where $pk_B$ is treated as the writer's identity. It can be seen that even though the forged $FI_{CB}=(I_{CB},CT_{CB})$ can pass the owner verification, it fails to pass the writer verification. Therefore, the ciphertext authentication protocol is able to defend against the forging attack.

\subsubsection{Defending against the disguise attack}
The defense against the disguise attack depends on that the verifier $B$ correctly recalls the first impression of prover $A$ in the challenge phase of the avatar authentication protocol. Assuming that there exists an adversary $\mathcal{C}$ can successfully initiate a disguise attack. It means that $\mathcal{C}$ is able to replace or forge a ciphertext based on $A$'s identity and write it into blockchain as $B$'s first impression of $\mathcal{C}$. According to the above analysis, it is impossible for $\mathcal{C}$ to implement these two attacks. Therefore, the avatar authentication protocol is able to defend against the disguise attack.

\begin{figure*}[htbp]
    \centering
     \subfigure[The VR Glass with an inserted iris camera.]{
     		\label{fig:IrisGlass}    \includegraphics[width=8.5cm]{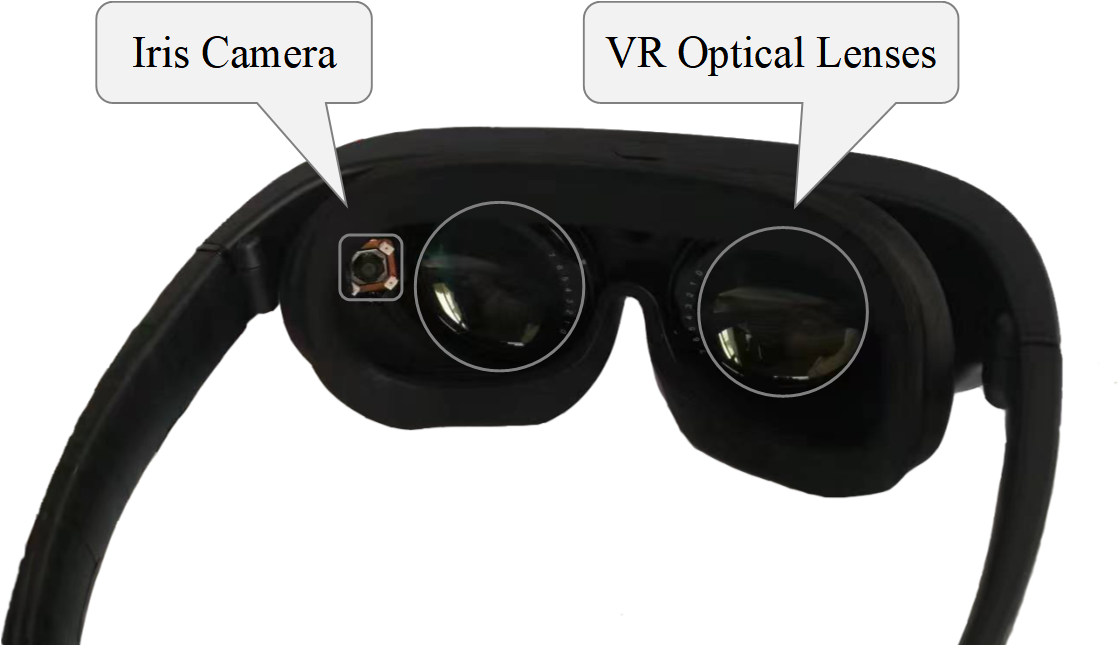}    }    \quad
    \subfigure[The avatar authentication system ]{
    		\label{fig:EvalAuth}    \includegraphics[width=8.5cm]{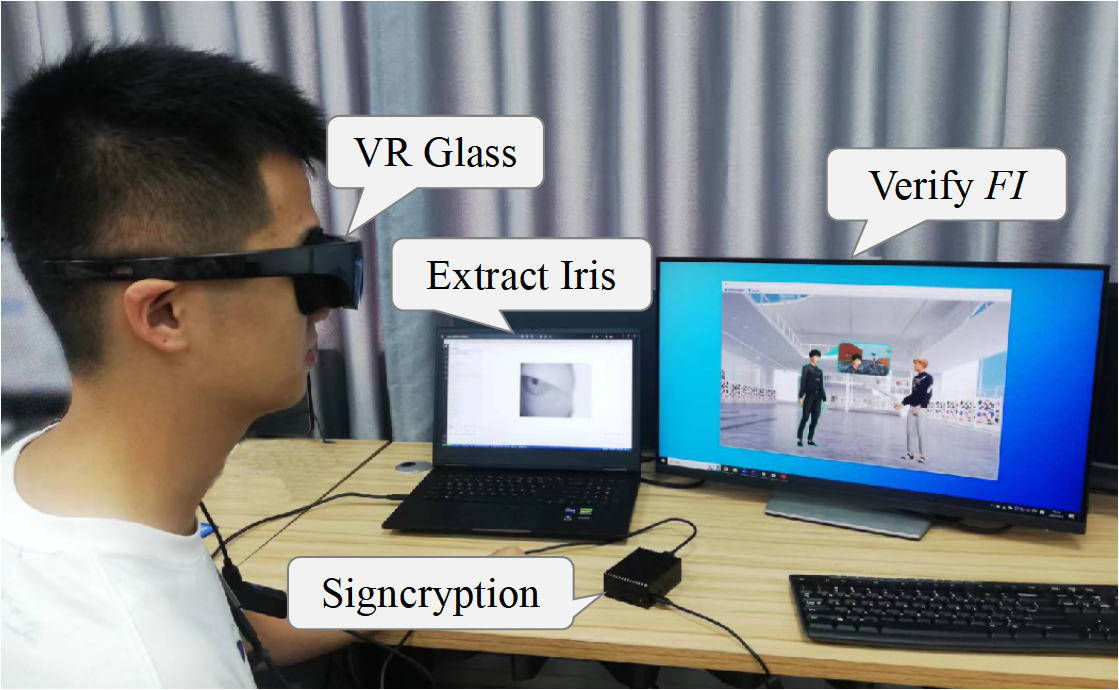}    }    \quad
    \caption{The modified VR Glass and the avatar authentication system, where avatar $A$ and avatar $B$ use the same device for authentication.}
   
\end{figure*}

\subsubsection{{Defending against the privacy-inference attack}}
{The adversary $C$ can reveal the privacy of user $A$ using two ways. The first is to infer the metaverse-meeting scene $Img_X$ of user $X$ based on $CT_{XA}$ implying in the first impression $FI_{XA}$. The other way is to reveal $A$'s identifiable information $Mid_A$ from $Hid_A$ implying in $MIT_A$. To infer $Img_X$ from $CT_{XA}$, the adversary $C$ should construct  $I_{XA}$ based on $Hid_X$ in $Avatar_X$ and $Hid_A$ in $Avatar_A$ to get $CT_{XA}$ from IPFS and construct a series of $\{I_{iA}=Hid_i||Hid_A\}_{i=1,2,...,n}$ based on $\{Hid_i \}_{i=1,...,n}$ to get $\{CT_{iA}\}_{i=1,2,..,n}$. Since the proposed signcryption scheme satisfies confidentiality, the adversary $C$ fails to gets any information of $Img_X$ from $CT_{XA}$ and $\{CT_{iA}\}_{i=1,2,...,n}$. On the other hand, it is difficult to infer $A$'s identifiable information $Mid_A$ from $Hid_A$. Assume that the adversary $C$ has obtained a series of $\{MIT_i\}_{i=1,...,n}$ from the blockchain and extracts the hased information $\{Hid_i\}_{i=1,...,n}$. Since $Hid_A$ and $\{Hid_i\}_{i=1,2,...,n}$ are both generated with a strongly secure hash algorithm, the adversary $C$ is unable to infer the identity information  $Mid_A$ based on $Hid_A$ and $\{Hid_i\}_{i=1,2,...,n}$. Therefore, the avatar authentication protocol can defend against the privacy-inference attack.}

\section{Performance Evaluation} \label{sec:PerfEval}

To evaluate the performance of the proposed protocol, we design a simplified metaverse platform.

\subsection{Experimental Setup}
The devices used by the current metaverse interaction include the head-mounted display (HMD), personal computer (PC), smartphone (SP), and low-computation-power device (LCPD). Since the current HMD does not support the iris authentication, we insert an iris camera into the HMD (HUAWEI VR Glass) as shown in Fig. \ref{fig:IrisGlass}. We simulate the low-computation-power device using the Raspberry Pi. The hardware parameters of devices are shown in TABLE \ref{tab:Device}.

\begin{table}[htbp]

      \centering
      \caption{Parameters of different devices }
      \label{tab:Device}
      \begin{threeparttable}
      \begin{tabular}{p{65 pt}<{\centering}p{75 pt}<{\centering}p{30 pt}<{\centering}p{25 pt}<{\centering}}
          \hline
        \specialrule{0em}{1pt}{1pt}
        Type & Device & CPU & ARM  \\
        \specialrule{0em}{1pt}{1pt}
           \hline
          \specialrule{0em}{1pt}{1pt}
          HMD & HUAWEI VR Glass  & --  & --  \\
          Desktop computer & DELL Precision 3650 & 2.8 GHz  & 64 GB  \\
          Laptop computer & HP OMEN 16 & 2.2 GHz  & 16 GB  \\
          Smart Phone & HUAWEI P40  & 2.8 GHz  & 8 GB  \\
          Raspberry Pi &  Raspberry Pi 4B & 1.5 GHz  & 4 GB \\
      \specialrule{0em}{1pt}{1pt}
        \hline
      \end{tabular}
\end{threeparttable}
\end{table}

{In the implementation, we use Java as the main programming language to build the authentication system. The IPFS is used to store source files such as $MIT$ and $CT$ and the blockchain platform FISCO BCOS 2.0 is used to store the extraction index returned from IPFS. The smart contracts are built through Solidity to realize the data access of IPFS and blockchain. The implementation of the authentication system is shown in Fig. \ref{fig:EvalAuth}, where the user $A$ acting as an adversary uses a laptop to interact with the user $B$ who manipulates his/her avatar using a desktop computer.}

{We use the avatar authentication protocol shown in Fig. \ref{fig:AvatarAuthProt} to introduce the implementation parts corresponding to the users' devices. The authentication protocol is implemented with four modules, namely the main protocol module, the iris extraction module, the signcryption module, and the verification module. The main protocol module, the iris extraction module, and the verification module are implemented on the metaverse platform, while the signcryption module is implemented on a trusted device. The iris extraction module wakes up the iris camera in the VR glass to capture eye images and extract iris samples. The functionalities of these modules are presented as follows. }

\begin{itemize}
\item {Main protocol module: This module serves as the dominant interface to call each sub-module for authenticating avatars. The module is implemented on the user's devices (e.g., the VR glass or the adhere computing device). }

\item {Iris extraction module: The steps ``$M^\prime \leftarrow$ Samples  biometrics” and “$M_{a}^{\prime}\leftarrow$ 
 Embeds challenge $C_{a}$” in the response phase of the protocol is realized in the iris extraction module, which is implemented by the user's device (e.g., the VR glass).}

\item  {Signcryption module: We realize this module according to the step “$CT_{A}^{\prime}\leftarrow SC(sk_{A}$, $h_{A},M_{a}^{\prime},pk_{B})$” in the response phase of the protocol and the step “$CT_{AB}\leftarrow SC(sk_{A},h_{A},Img_{A}$, $pk_{B})$” when creating the $FI_{AB}$, which is implemented on the user's device (e.g., the VR glass or the adhere computing device).}

\item {Verification module: The signatures and iris samples are verified on the user's device (e.g., the VR glass or the adhere computing device).}

\end{itemize}

\begin{table*}[htbp]
\footnotesize
 \centering
      \caption{The successful rate of the disguising attack}
      \label{tab:DiguiseAttack}
      \begin{threeparttable}
      \begin{tabular}{p{60 pt}<{\centering}p{80 pt}<{\centering}p{150 pt}<{\centering}p{150 pt}<{\centering}}
          \hline
        \specialrule{0em}{1pt}{1pt}
       Group &  Authentication factor &      Rate of providing private information & Rate of performing designated actions  \\
        \specialrule{0em}{1pt}{1pt}
           \hline
          \specialrule{0em}{1pt}{1pt}
            $Group_1$  &  - &   50\% & 40\% \\
            $Group_2$  & Name &  50\% & 50\% \\
            $Group_3$  & $ID$ &     30\% & 20\% \\
            $Group_4$  &   $FI$ &   0 & 0 \\
      \specialrule{0em}{1pt}{1pt}
        \hline
      \end{tabular}
       \begin{tablenotes}
        \footnotesize
         \item[] ``$-$''  means no information displayed.
      \end{tablenotes}
\end{threeparttable}
\end{table*}

\subsection{Evaluation metrics and comparative benchmarks}
{
The evaluation metrics are classified into attack metrics and performance metrics. The attack metrics are: 
\begin{itemize}
  \item Rate of providing private information: The proportion of the tested persons who share his/her private information with the ``friend" (i.e., malicious user).
  \item Rate of performing designated actions: The proportion of the tested persons who perform the designated actions by the ``friend" (i.e., malicious user).
\end{itemize}

The performance metrics are:
\begin{itemize}
  \item Execution Time of Signcryption: The time required to run the signcryption for constructing the authentication factors and computing the verification process.
  \item Cost of IPFS and Blockchain: 
   \begin{itemize}
     \item  The time consumption for the operations over IPFS and blockchain.
     \item The storage consumption for the operations over IPFS and blockchain.
   \end{itemize}
  \item Execution Time of Authentication Protocol:
     \begin{itemize}
     \item  The time required for executing the key steps of the avatar authentication protocol.
     \item The time required for executing the key steps and full steps of the ciphertext authentication protocol.
     \item The overall time required for carrying out the metaverse interaction for the authentication. 
   \end{itemize}
 \end{itemize}

The comparative benchmarks are also divided into the attack benchmarks and the performance benchmarks.

The attack benchmarks are:
\begin{itemize}
  \item There is no authentication factor for the avatar.
  \item Only the name is used to authenticate the avatar.
  \item Only the ID number is used to authenticate the avatar.
\end{itemize}

The performance benchmarks are:
\begin{itemize}
  \item The signcryption scheme proposed in Li's work \cite{Li2007EffiSC}.
  \item The avatar authentication protocol proposed in Yang's work \cite{Yang2023Trace}.
\end{itemize}
}

\subsection{{Evaluation of the Disguise Attack}}
{We select four users as victims to construct four control groups $\{Group_i\}_{i=1,...,4}$, in which each group contains ten friends. An attack environment is designed by combining a voice plugin and our simplified metaverse platform. During the attack, the adversary trains a voice model for each victim based on the voice plugin and then talks with ten friends of the victim using the voice model. The attack has two goals. One is to deceive the friends to disclose his/her private information such as ID card number. The other is to induce the friends to perform designated actions. The successful rate of the attack is shown in TABLE \ref{tab:DiguiseAttack}.}

{From the table, for the friends in $Group_1$ who display nothing while using the public keys and iris templates as the authentication factors, the probability that the adversary successfully obtains the user's private information and induces the users to perform designated actions is 50\% and 40\%, respectively. For the friends in $Group_2$, they use the name as the authentication factor. The successful rate of the disguising attack is 50\% for providing private information and performing designated actions. For the friends in $Group_3$ with $ID$ as the authentication factor, the probability that the adversary successfully obtains the user's private information and induces the users to perform designated actions is 30\% and 20\%, respectively. For the users in $Group_4$, they use $FI$ as an auxiliary authentication factor. The successful rate of the disguising attack is zero. Therefore, the proposed authentication method can effectively defend against the disguise attack.}

\subsection{Execution Time of Signcryption} 
\begin{figure}[htbp]
\begin{center}
    \includegraphics[width=0.47\textwidth]{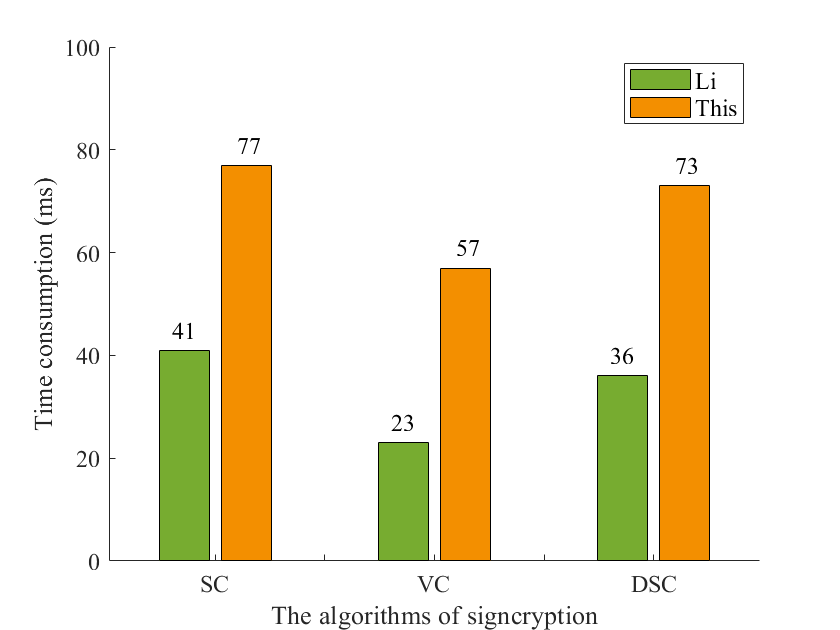}
    \caption{\small{The execution time of the signcryption scheme.}}
    \label{fig:SignCryptTime}
    \vspace{-0.5em}
 \end{center}
\end{figure}

To analyze the execution time of signcryption, we implement the proposed scheme and Li's scheme \cite{Li2007EffiSC} with Java codes. During the evaluation, we perform the algorithms of signcryption on the desktop computer and record the execution time of these two schemes. It can be seen from Fig.\ref{fig:SignCryptTime} that the execution time of the proposed signcryption is larger than that of Li's scheme (about 30ms). However, the proposed signcryption scheme achieves public verifiability using the chameleon signature. Overall, the execution time of the proposed signcryption scheme is less than 80ms, which does not affect the real-world use. Therefore, the proposed signcryption scheme meets the application requirement for metaverse users.

\subsection{Cost of IPFS and Blockchain}\label{sec:signcypexe} 

{This part evaluates the cost of $MIT$ and $FI$ operated on blockchain and IPFS, including time consumption and storage consumption.}

\begin{figure*}[htbp]
    \centering
     \subfigure[The time consumption of $MIT$ and $FI$]{
     		\label{fig:BlockTime}    \includegraphics[width=8.5cm]{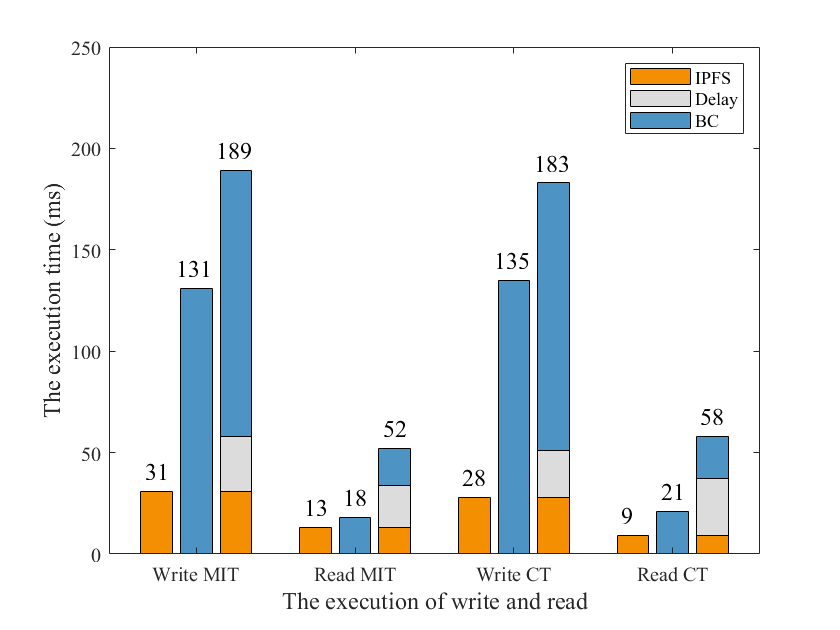}    }    \quad
    \subfigure[The storage consumption of $MIT$ and $FI$]{
    		\label{fig:StoreCost}    \includegraphics[width=8.5cm]{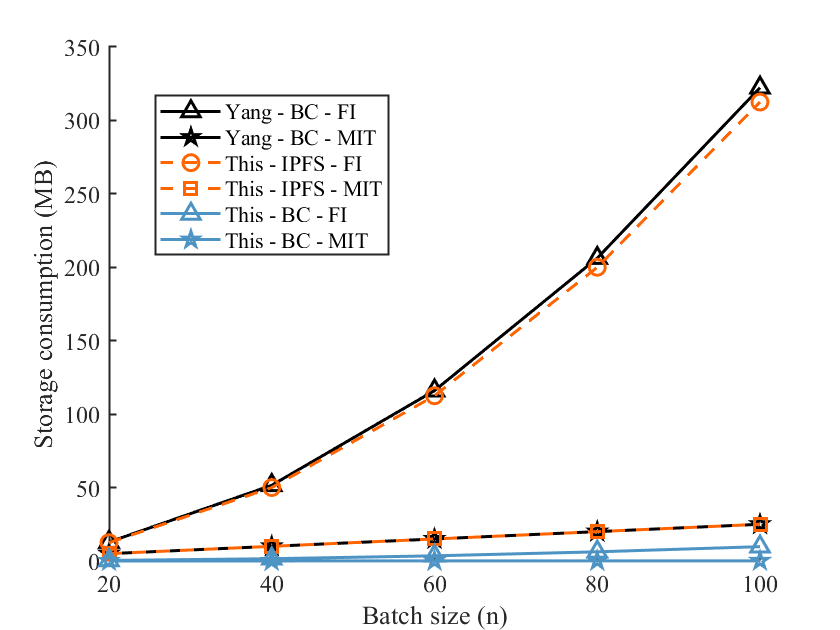}    }    \quad
    \caption{
 The consumption on IPFS and blockchain (i.e., BC), which includes the consumption of $MIT$ and $CT$ operated on IPFS and BC. The ``Delay'' means the interaction delay between IPFS and BC.}
\end{figure*}

{Here we evaluate the time consumption for the $MIT$ and $FI$ operated on IPFS and blockchain, respectively. For $MIT$, we first write $MIT$ to IPFS and then write the returned file index $FID$ to the blockchain as a transaction. For $FI$, we first write the $CT_{AB}$ about $FI_{AB}$ to IPFS and then write the extraction index $I_{AB} =(Hid_A||Hid_B, FID_{AB})$ to the blockchain as a transaction. The experimental results are shown in Fig.\ref{fig:BlockTime}. We can see that the proposed approach writes a $MIT$ to IPFS and blockchain is about 30ms and 130ms, respectively, and the total consumption is about 190ms. For reading a $MIT$, the time consumption is about 50ms. The time consumption for writing is less than 200ms, which is fast enough to satisfy the requirement for the avatar's authentication. Besides, since both IPFS and FISCO BCOS 2.0 support large-scale data in storage and access, the proposed system can support large-scale dynamic authentication in the metaverse.}

\textbf{Storage Consumption of \textit{MIT} and \textit{FI} :} Traditional authentication methods \cite{Ryu2022MutMeta, Yang2023Trace} store the user's identity parameters on the blockchain to achieve mutual authentication between avatars. This work combines with IPFS to reduce the storage consumption on the blockchain. To analyze the consumption of the proposed authentication protocol, we conduct extensive experiments to compute the storage consumption of $MIT$ and $FI$. For {$MIT=(SN$, $Hid,pk,T)$}, since the iris template $T$ is a BMP image of 193 kb, we prepare 256 kb storage space for each $MIT$. As for $FI=(I,CT)$, since the $I=$ {$Hid_A||Hid_B$} contains two $Mid$ of 1024 bit and the $CT$ contains a JPG image of 22.4 kb, we prepare 1 kb and 32 kb storage space for each $I$ and $CT$, respectively. In the experiments, we set the number of friends of each user to 20, 40, 60, 80, and 100. If all friends need to do authentication, the storage consumption on blockchain and IPFS for the user is shown in Fig.\ref{fig:StoreCost}. In terms of blockchain, the consumption of $FI$ in Yang's method \cite{Yang2023Trace} increases quadratically, while this work increases linearly. Overall, Yang's consumption of $FI$  and $MIT$ are significantly higher than this work. Therefore, the proposed avatar authentication protocol achieves anti-disguise with a low storage consumption on the blockchain compared to Yang's method.

\begin{figure*}[htbp]
    \centering
     \subfigure[Avatar authentication protocol]{
     		\label{fig:AvatarAuthTime}    \includegraphics[width=8.5cm]{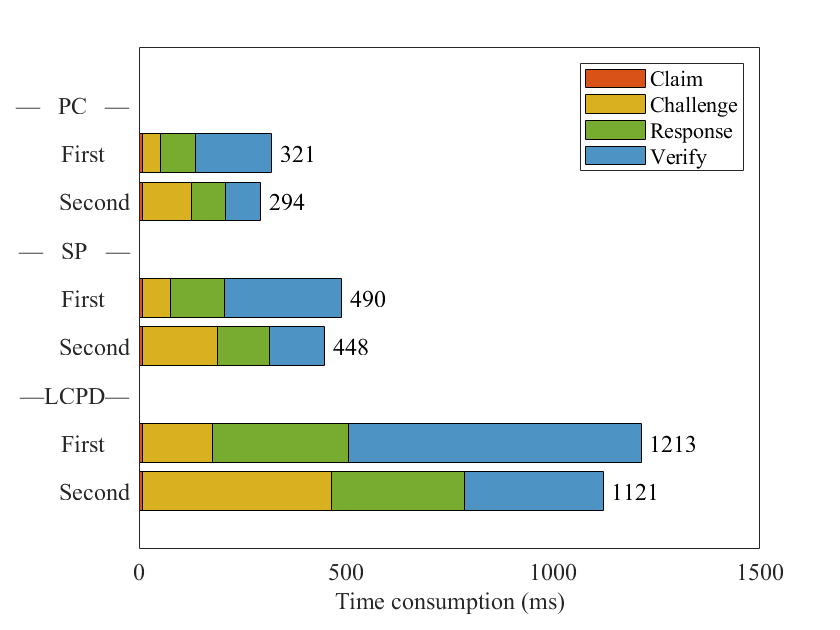}    }    \quad
    \subfigure[Ciphertext authentication protocol]{
    		\label{fig:CryptAuthTime}    \includegraphics[width=8.5cm]{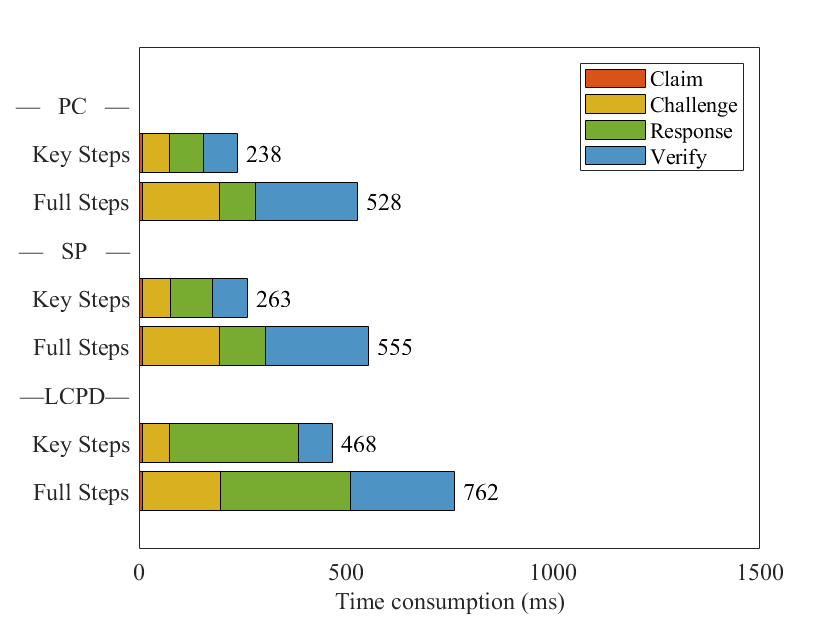}    }    \quad
    \caption{The execution time of the avatar authentication protocol and the ciphertext authentication protocol on different platforms.}
    \vspace{-1.5em}
\end{figure*}

\begin{figure*}[htbp]
    \centering
     \subfigure[The interaction time of the first meeting]{
     		\label{fig:TotalCostFirst}    \includegraphics[width=8.5cm]{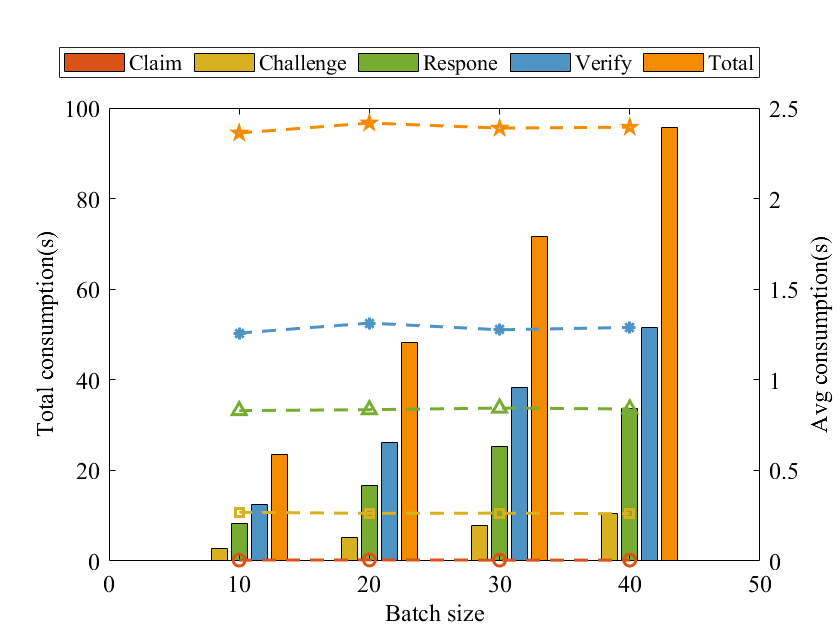}    }    \quad
    \subfigure[The interaction time of the second meeting]{
    		\label{fig:TotalCostSec}    \includegraphics[width=8.5cm]{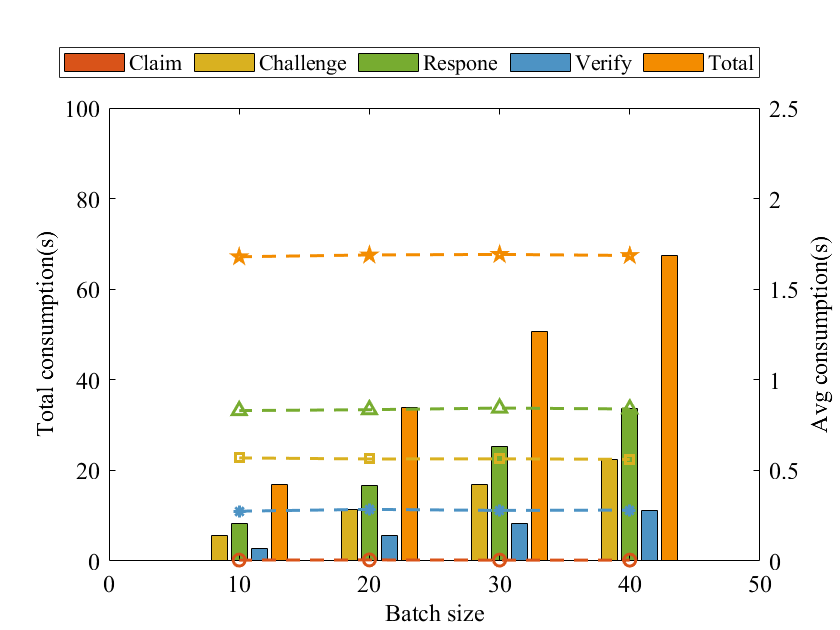}    }    \quad
    \caption{The interaction time of the authentication for the first and the second encounters, respectively.}
    \vspace{-1.0em}
\end{figure*}

\subsection{Execution Time of Authentication Protocol} \label{sec:authenkey}
In this part, we analyze the execution cost of the proposed authentication protocols, which involves time cost and storage consumption.

\textbf{Authentication Time of Key Steps of the Avatar Authentication Protocol}: To analyze the execution time of key steps in the avatar authentication protocol, we take the first and second metaverse meetings as the primary authentication scenarios. The algorithms $SC, VC$, and $DSC$ in the avatar authentication protocol are treated as key steps. In the claim phase, since it only involves the submission of identity parameters while the above three algorithms are not needed, the execution time of the key step is zero. In the challenge phase, we take the verification on $VID_{A}$ and $FI_{AB}$ as key steps. In the response phase, we take the generation of $CT_A^\prime$ as the key step. In the verify phase, we take the de-signcryption of $CT_A^\prime$ as the key step. In the experiments, the avatars $A$ and $B$ use the same device for authentication. The device can be a laptop computer (represented as ``PC''), smartphone (represented as ``SP''), and low-computation-power device like Raspberry Pi (represented as ``LCPD''). The authentication time of key steps {when avatars meet for the first and second time} is shown in Fig.\ref{fig:AvatarAuthTime}. It can be seen that the overall execution time is about 500 ms on the platforms of the PC and smartphone, which has little impact on the user experience. However, the overall execution time on the low-computation-power device is more than 1000ms, which might not be applicable in this case and indicates that sometimes there should be a trade-off between security and device cost. In general, the proposed signcryption scheme is the potential to be used in the PC and smart phone platforms to realize anti-disguise authentication.

\textbf{Ciphertext Authentication Time of Key Steps and Full Steps:}
To analyze the authentication time on ciphertext, we use the IPFS platform to store the ciphertext of the first impression and utilize the blockchain FISCO BCOS 2.0 to store the index. In experiments, we treat the algorithms for signcryption as the key steps and the complete protocol as full steps. The execution time of ciphertext authentication is shown in Fig.\ref{fig:CryptAuthTime}. It can be seen that the execution time of key steps and full steps on all platforms is less than 600 ms, which indicates that the storage and verification of the first impression on all platforms is very efficient. Therefore, the proposed ciphertext authentication protocol meets the real-world requirements for both efficiency and security requirements under the replacing and forging attacks.

\textbf{Overall Execution Time of the Metaverse Interaction for the Authentication:} To evaluate the overall execution time of the metaverse interaction, we manipulate avatar $A$ through a HUAWEI VR Glass, a laptop computer, and a Raspberry Pi as shown in Fig.\ref{fig:EvalAuth}, where the VR Glass is used to display the interaction scenario, the laptop computer is used to extract the iris image, and the Raspberry Pi is used to generate the ciphertext of the first impression. We manipulate avatar $B$ through a desktop computer, which is used to verify the first impression of $A$. For generating the ciphertexts of the iris image and first impression, we first use a 128-bit symmetric key to encrypt the iris image and the first metaverse-meeting scene and then utilize the proposed signcryption mechanism to encrypt the symmetric key. During the interaction, the times of authentication are set as 10, 20, 30, and 40, respectively, to obtain an average authentication time. The execution time of the interaction for the avatar's first meeting and second meeting are shown in Fig.\ref{fig:TotalCostFirst} and Fig.\ref{fig:TotalCostSec}, respectively. From Fig.\ref{fig:TotalCostFirst}, the total time for the avatar's first meeting is about 2.5s, in which the most time-consuming process is the verification phase. It seems longer than that of the second meeting. The reason is that, during the first meeting, the avatars should create the first impression. For the second meeting, from Fig.\ref{fig:TotalCostSec}, the total interaction time is reduced to around 1.6s. The overall interaction time for the authentication does not affect the participation of users in their main applications/services.

\section{Related work} \label{sec:RelaWork}
The metaverse is still in its infancy and lacks feasible authentication methods to avoid disguise attacks. In this section, we sort out the traditional authentication methods in the metaverse and signcryption mechanisms to provide references for designing practical authentication schemes with anti-disguise.

\begin{table*}[htbp]
    \centering           
     \caption{ The Advantages of Different Schemes}
      \label{tab:Advantages}
      \begin{threeparttable}      \begin{tabular}{p{70 pt}<{\centering}p{80 pt}<{\centering}p{80 pt}<{\centering}p{70 pt}<{\centering}p{50 pt}<{\centering}p{70 pt}<{\centering}}
        \hline
        \specialrule{0em}{1pt}{1pt}
         Schemes& $Scenes$ & $Factors$ &  $Decentralization$ & $Mutual$  &   $Anti\text{-}disguise$\\
        \hline 
        \specialrule{0em}{1pt}{1pt}
      Mathis \cite{Mathis2020Rubik} & login &  password \& behavior &  \ding{56} & \ding{56} &  \ding{56}  \\
        Wang \cite{Wang2022Egocentric} & login & iris &  \ding{56} & \ding{56} &\ding{56}  \\
        Sethuraman \cite{Sethuraman2022Metakey} & login & key \& face & \ding{56} & \ding{52} &\ding{56}  \\       
      Ryu \cite{Ryu2022MutMeta} & login \& interaction & key \& fingerprint &  \ding{56} & \ding{52} &  \ding{56}\\
        Kim \cite{Kim2023Idtifier} & login \& interaction & key \& fingerprint &  \ding{56} & \ding{52} & \ding{56}\\
      Patwe  \cite{Patwe2023User}  & login \& interaction & password &  \ding{52} & \ding{52} &  \ding{56}\\    
     Zhong  \cite{Zhong2023CAN}  & login \& interaction & ear &  \ding{52} & \ding{52} & \ding{56}\\   
     Yang \cite{Yang2023Trace}  & interaction & key \& iris  & \ding{52} & \ding{52} &  \ding{56}\\
    This work & interaction & key \& iris \& FI & \ding{52} & \ding{52} & \ding{52}\\
      \hline
      \end{tabular}    
\end{threeparttable}
\end{table*}

\subsection{Metaverse Authentication}
Avatar authentication methods mainly involve two types to guarantee the security of the user's identity, which are login authentication between avatars and platforms \cite{Mathis2020Rubik,Wang2022Egocentric,Sethuraman2022Metakey} as well as interaction authentication between avatars and avatars \cite{Ryu2022MutMeta,Yang2023Trace}.

For the methods of login authentication, platforms verify the avatar's identity based on biometrics. Mathis \textit{et al.} \cite{Mathis2020Rubik} combined the user's password and biological behavior characteristics to build a multi-factor authentication model. In this model, while the user enters the login password in the Rubik's cube, the VR device continuously collects user behavior to achieve two-factor authentication. {Aiming at the close fit between VR glasses and human eyes, Wang \textit{et al.}  \cite{Wang2022Egocentric} constructed a metaverse user identification method based on iris features, which is applicable to various metaverse scenarios, such as login and payment.} Sethuraman \textit{et al.}  \cite{Sethuraman2022Metakey}  proposed a novel and seamless passwordless multi-factor authentication system {based on built-in key and face trait}  to provide passwordless authentication with the high security of biometric recognition at login. 

{The interactive authentication mainly utilizes biometric features and signature keys to realize the authentication between avatars and avatars. To realize the login authentication and interactive authentication, Ryu \textit{et al.}  \cite{Ryu2022MutMeta}  combined fingerprint biometrics and elliptic curves to construct a secure mutual multi-factor authentication scheme. The privacy issue in Ryu's scheme was addressed by Kim \textit{et al.}   \cite{Kim2023Idtifier}, which designed a multiple-factor authentication framework based on decentralized identifiers to guarantee the avatar's anonymity. Considering a malicious server, Patwe \textit{et al.} \cite{Patwe2023User} proposed a blockchain-based authentication architecture to address the server spoofing attack and the identity interoperability issue. To achieve continuous authentication during the interactions, Zhong \textit{et al.}  \cite{Zhong2023CAN}  constructed a continuous, active, and non-intrusive multi-factor authentication scheme based on deep learning, which uses the audio as the challenge and the reflected sound as the responses during the interaction process. Yang \textit{et al.}  \cite{Yang2023Trace} constructed a two-factor authentication framework based on the user's iris and keys to guarantee virtual-physical traceability. There are many related studies on multi-factor authentication in other fields but they have the same characteristic in that they make use of multiple identity information to guarantee security and efficiency during the authentication procedure \cite{Tahir2023LightMulti,Braeken2023RealTime,Han2024Enhanced}.}

{For mutual authentication, the above authentication schemes treat the user's device, such as HMD, as a trusted unit to authenticate the manipulator. However, they fail to defend against the disguise attack. That is, the authentication approaches cannot prevent the adversary from manipulating an avatar to carry out intrusive activities with his/her own device. Therefore, in this paper, we use more factors to authenticate the interaction process and create the first impression to alleviate the user's burden of remembering ID numbers. The technical difference is shown in TABLE \ref{tab:Advantages}. The proposed approach can achieve decentralized and mutual authentication and has the capability of defending against the disguise attack.}

\subsection{Signcryption}

The first impression is a more practical method for ensuring anti-disguise. Signcryption \cite{Zheng1997Signcryption} is a cryptographic primitive targeting to provide conﬁdentiality and unforgeability simultaneously, which is able to guarantee the privacy and verifiability of first impressions. At present, signcryption is widely applied in the Industrial Internet of Things (IIoT), Internet of Vehicles (IoV), Internet of Healthcare Things (IoHT), and related fields.

In the field of IIoT, Xiong \textit{et al.}  \cite{Xiong2021Test} proposed a heterogeneous signcryption scheme allowing the delegated cloud server to execute equality tests on ciphertexts. Facing untrusted entities from leaking user privacy, Chen \textit{et al.}  \cite{Chen2022Offline}  constructed an improved certiﬁcateless online/ofﬂine signcryption scheme to achieve lower computational overhead. Dohare \textit{et al.}  \cite{Dohare2022CLASS} proposed a proﬁcient certiﬁcateless aggregated signcryption scheme, which provides a data aggregation element to achieve data authentication. In the field of IoV, Ali \textit{et al.}  \cite{Ali2022ECCHSC} proposed an elliptic curve cryptosystem-based hybrid signcryption protocol that satisﬁes the security requirements for heterogeneous vehicle-to-infrastructure communications.  Ullah \textit{et al.}   \cite{Ullah2023Condi} proposed a conditional privacy-preserving heterogeneous signcryption scheme for IoV to relieve the problem of high computational overhead. In the field of IoHT, Xiong \textit{et al.}  \cite{Xiong2022WBAN} proposed a heterogeneous signcryption scheme from an identity-based cryptosystem to public key infrastructure with an equality test for WBANs.

Although these schemes solve
various security issues in the fields of IIoT, IoT, and IoHT,
the identities implied in ciphertexts need to be decrypted
to verify, which is not suitable for non-decryption scenarios
of the first impression. Therefore, it is urgent to construct
a signcryption scheme with public verifiability to support
ciphertext authentication.

\subsection{Digital watermarking}
{Digital watermarking is a technique that embeds identifiable information into the digital carrier. It is generally used to verify the ownership of user-generated data. Wang \textit{et al.} [22] constructed a federated learning framework based on digital watermarking in the social metaverse, enhancing the privacy-utility trade-off and supporting the model ownership verification to defend against AI model thefts. If the identifiable information of an avatar is embedded into the carrier of the first impression, the verifiability of the first impression can be guaranteed. However, it is worth noting that, the watermarked first impression (i.e., an image of the first-meeting scene) looks the same as that has no watermarking. If the watermarked first impression is stored on the blockchain, the user's private social information can be inferred. Therefore, we need to develop a new method to guarantee the privacy and verifiability of the first impression. Encryption is a common technique to protect the user's privacy. However, because the existing encryption scheme fails to simultaneously guarantee the privacy and verifiability, we propose a chameleon signcryption mechanism to realize this privacy-protection goal.}

\section{Conclusion and Future Work } \label{sec:Conclu}
In the metaverse, the disguise attack is a kind of appearance deception attack in which an adversary generates the same avatar as a target one to mimic others. To defend against this attack, this paper proposed a multi-factor avatar authentication protocol by combining the ``first impression'' generated with the metaverse scenes, realizing the merge of digital and unconscious verification of the avatar's core identity. To prevent the replacing and forging attacks on the first impression, we proposed a chameleon signcryption mechanism with public verifiability and designed a ciphertext authentication protocol, which guarantees the consistency between the writer's and the owner's identities implying in a ciphertext of first impression. Overall, the proposed protocol has the capability of defending against the disguised attack in an effective and efficient way. 

This work only considers the authentication issue of human-driven avatars. In the future metaverse, however, there might have many AI-driven avatars interacting with human-driven avatars to provide users with services that break through physical limitations. Since the metaverse is an open digital ecosystem, any developer can deploy his/her own AI avatars in the metaverse. In that case, the malicious users can build AI-driven avatars in the metaverse to defraud users' private information and financial issues. Therefore, it is necessary to design an authentication mechanism to inhibit attackers from deceiving other users through AI-driven avatars.

\ifCLASSOPTIONcaptionsoff
  \newpage
\fi

\section{Security proof for the chameleon signcryption}\label{sec:AppProof}

This part proves the security of the chameleon signcryption mechanism, including SC-EUF-CMA, C-EUF-CMA, and SC-IND-CCA.

\subsection{SC-EUF-CMA} \label{sec:ProofScEuf}

\textbf{Theorem 1}. If the DCDH assumption holds on $\mathbb{G}$, the proposed signcryption mechanize satisfies SC-EUF-CMA.

\begin{figure}[htbp]
  \begin{center}
        \fbox{\parbox{8.5cm}{
        \begin{flushleft}
          \hspace{0.5em}$\textbf{Experiment} \quad Exp_{CH\text{-}SC,\mathcal{A}}^{SC\text{-}EUF}(\mathcal{K})$\\
          \hspace{0.5em}$Parms \leftarrow Setup(\mathcal{K});$\\
          \hspace{0.5em}$(pk,sk) \leftarrow KeyGen(Parms)$;\\
          \hspace{0.5em}$L_1\leftarrow \varnothing,L_2\leftarrow \varnothing,L_3\leftarrow \varnothing,L_s \leftarrow \varnothing, c \stackrel{R}{\leftarrow} \mathbb{Z}_q;$\\
          \hspace{0.5em}$(h,R) \leftarrow Hash(pk,M), L_1 \leftarrow (M,t,m);$\\
          \hspace{0.5em}$(M^*,CT^*,pk^*) \leftarrow \mathcal{A}^{\mathcal{H}_{i\in \{1,2,3\}}(\cdot),\mathcal{SC}(\cdot)}(pk,h) $; \\
          \hspace{2.0em}$m^\prime \leftarrow \mathcal{H}_1(M^\prime), \; L_1 \leftarrow (M^\prime,t^\prime,m^\prime)$; \\
          \hspace{2.0em}$w^\prime \leftarrow \mathcal{H}_2(K^\prime,y^\prime,Y^\prime), \; L_2\leftarrow(K^\prime,y^\prime,Y^\prime, w^\prime)  $; \\        
          \hspace{2.0em}$M^{\prime\prime} \leftarrow \mathcal{H}_3(K^\prime,Z^\prime,y^\prime), \; L_3\leftarrow(K^\prime,Z^\prime, y^\prime, M^{\prime\prime})  $; \\        
          \hspace{2.0em}$CT^\prime \leftarrow \mathcal{SC}(M^\prime,pk^\prime), \; L_s\leftarrow (M^\prime,K^\prime,Z^\prime,R^{\prime\prime},pk^\prime)$; \\
          \hspace{0.5em}if $DSC(pk,h,CT^*, sk^* )=M^*\bigwedge (M^*,CT^*)\notin L_s$, \\
          \hspace{2.0em}return 1;\\
          \hspace{0.5em}else return 0.\\
        \end{flushleft}
        }}

    \caption{The game model of SC-EUF-CMA.}
    \label{fig:SC-EUF-CMA}
    \vspace{-1.5em}
    \end{center}
\end{figure}

\textbf{Proof 1.}  For contradiction, if there exists a polytime adversary $\mathcal{A}$ who wins the game $Exp_{CH\text{-}SC,\mathcal{A}}^{SC-EUF}(\mathcal{K})$ with a non-negligible advantage $\epsilon$, there exists an adversary $\mathcal{B}$ who can solve the DCDH problem with a non-negligible advantage $\epsilon^\prime$.

Suppose $\mathcal{B}$ is given a random instance  $(g, g^a, g^b) \in \mathbb{G}^3$ of the DCDH problem. Based on the instance, $\mathcal{B}$ sets up a simulated environment of SC-EUF-CMA as $Exp_{CH\text{-}SC,\mathcal{A}}^{SC\text{-}EUF}(\mathcal{K})$ for $\mathcal{A}$ and runs $\mathcal{A}$ as a subroutine to ﬁnd the solution $g^{b/a}$.

The simulation $Exp_{CH\text{-}SC,\mathcal{A}}^{SC\text{-}EUF}(\mathcal{K})$ is divided into three phases including initialization, inquiry, and forgery. The details are as follows:

1) Initialization. The challenger $\mathcal{B}$ sets the public key $pk$ for the challenge and chameleon tuple $(h, M, R)$ by these four steps: (i) $\mathcal{B}$ constructs the system parameter $Parms$ as \ref{sec:CHSC}, (ii) sets the challenge key $pk=y=g^a$, (iii) initializes the auxiliary lists $L_1,L_2,L_3,L_s$ for recording the queries on oracles $\mathcal {H}_1,\mathcal{H}_2,\mathcal{H}_3$, $\mathcal{SC}$, where $L_1=(M^\prime,t^\prime,m^\prime)$, $L_2=(K^\prime,Z^\prime,M^{\prime\prime})$, $L_3=(K^\prime,y^\prime,Y^\prime,w^\prime)$, $L_s=(M^\prime, K^\prime,Z^\prime,R^{\prime\prime},pk^\prime)$; (iv) generates an initial chameleon tuple $(h,M,R)$ and records the entry $ (M,t,m)$ in $L_1$, where $c\stackrel{R}{\leftarrow} \mathbb{Z}_q$, $h=g^c$, $t\stackrel{R}{\leftarrow} \mathbb{Z}_q$, $m=\frac{ g^cy^t}{y^c}$, $R=\frac{g^c}{g^t}$.

2) Inquiry. Before inquiring, $\mathcal{B}$ assumes that $\mathcal{A}$ will forge a signcryption on the $j$th result of inquiring  $\mathcal{H}_1$. In the inquiry process, $\mathcal{A}$  is able to inquire the oracle $\mathcal{H}_1,\mathcal{H}_2,\mathcal{H}_3$, and $\mathcal{SC}$ for a certain number of times. $\mathcal{B}$ first returns $\mathcal{A}$ the corresponding answers $m^\prime,w^\prime,M^{\prime\prime}$, and $CT^\prime$, then records parameters using the corresponding lists $L_1,L_2,L_3$, and $L_s$.  

——$\mathcal{H}_1(M^\prime) \rightarrow m^\prime$. $\mathcal{B}$ retrieves $L_1$ by $M^\prime$ and returns $m^\prime \leftarrow L_1(M^\prime)$ upon receiving a query on $\mathcal{H}_1$ with $M^\prime$. If there is no corresponding entry, $\mathcal{B}$ randomly selects $t^\prime=t_i^\prime\stackrel{R}{\leftarrow} \mathbb{Z}_q$, calculates $m^\prime=m_i^\prime = \frac{g^cy^{t_i^\prime}} {y^c}$, and adds a new entry $(M^\prime,t^\prime,m^\prime)$ to $L_1$. It is worth noting that if $i=j$, $\mathcal{B}$ updates $ m^*=m_j^\prime=\frac{g^cy^{t_j^{\prime}}}{g^by^c}$, which is used to construct the solution for the DCDH problem. 

——$\mathcal{H}_2(K^\prime,y^\prime,Y^\prime)\rightarrow w^{\prime}$. $\mathcal{B}$ returns $w^{\prime} \leftarrow L_2(K^\prime,y^\prime,Y^\prime)$  upon receiving a query on  $\mathcal{H}_2$ with $(K^\prime,y^\prime,Y^\prime)$. If there is no corresponding entry, $\mathcal{B}$ randomly selects $w^\prime\stackrel{R}{\leftarrow}\{0,1\}^{n+l}$ and adds a new entry $( K^\prime, y^\prime, Y^\prime, w^\prime)$ to $L_2$. 

——$\mathcal{H}_3(K^\prime,Z^\prime, y^\prime)\rightarrow M^{\prime\prime}$.   $\mathcal{B}$ returns $M^{\prime\prime} \leftarrow L_3(K^\prime,Z^\prime, y^\prime)$  upon receiving a query on $\mathcal{H}_3$ with $(K^\prime,Z^\prime, y^\prime)$. If there is no corresponding entry, $\mathcal{B}$ randomly selects $M^{\prime\prime}\stackrel{R}{\leftarrow}\{0,1\}^n$ and adds a new entry $( K^\prime, Z^\prime, y^\prime, M^{\prime\prime})$ to $L_3$.

——$\mathcal{SC}(M^\prime, pk^\prime) \rightarrow CT^\prime$. Upon receiving a query on $\mathcal{SC}$ with $(M^\prime,pk^\prime=y^\prime)$, $\mathcal{B}$ retrieves $i\leftarrow L_1(M^\prime)$ such that $M_i^\prime=M^\prime$. To generate the corresponding signcryption, $\mathcal{B}$ randomly selects $k^\prime \stackrel{R}{\leftarrow} \mathbb{Z}_q$, calculates $K^\prime=g^{k^\prime}$, gets $m_i^\prime \leftarrow L_1(M_i^\prime) $, $t_i^\prime \leftarrow L_1(M_i^\prime)$, and calculates $R_i^\prime$ as 

 $\cdot$ if $i\neq j$, then $R_i^{\prime}=\frac{g^c}{g^{t_i^{\prime}}}$;

 $\cdot$ if $i=j$, then $\mathcal{B}$  stops and returns ``$\perp$'' indicating an error.

 $\mathcal{B}$ returns $CT=(K^\prime,Z^\prime,R^{\prime\prime})$  and adds a new entry $(M^\prime,K^\prime,Z^\prime,R ^{\prime\prime},y^\prime)$ to $L_s$ as follows:

 $Y^\prime=(y^\prime)^{k^\prime}$, $w^\prime \leftarrow \mathcal{H}_2(K^ \prime,y^\prime,Y^\prime)$, $Z^\prime=(M_i^\prime||R_i^\prime)\oplus w^\prime$, $M^{\prime\prime}\leftarrow \mathcal{H}_3(K^\prime,Z^\prime,y^\prime)$, $t^{\prime\prime}\leftarrow L_1(M^{\prime\prime})$, $R ^{\prime\prime}=\frac{g^c}{g^{t^{\prime\prime}}}$ . 

3) Forgery. $\mathcal{A}$ outputs $M^*$, $pk^*$ and $CT^*=(K^*,Z^*,R^*)$. If $DSC(pk,CT^*, h,sk^*)= M^*$ and $(M^*,CT^*) \notin L_s$, then  $\mathcal{A}$ wins.

In the simulation $Exp_{CH\text{-}SC,\mathcal{A}}^{SC\text{-}EUF}(\mathcal{K})$, if the guess \emph{j} from $\mathcal{B}$ is correct and $\mathcal{A}$ outputs a correct forgery,  $\mathcal{B}$ is able to output the solution $g^{b/a}$ of the DCDH problem based on $CT^*=(K^*,Z^*,R^*)$ as follows:

 $\mathcal{B}$  retrieves $t_j^\prime \leftarrow L_1(M^*)$ and $(K^*,y^*,Y_i^\prime,w_i^\prime
 )_{i \in [1,q_h]}\leftarrow L_2(K^*,y^*=pk^*)$, finds $w^*=w_i^\prime\leftarrow(K^*$, $y^*,Y_i^\prime,w_i^\prime)_{i \in [1,q_h]}$ such that $e(K^*,y^*)=e(Y_i^\prime,g)$, calculates $R^\prime=Z^*\oplus w^*$, and outputs $g^{(b/a)}=R^\prime \cdot \frac{g^{t_j^{\prime}}} {g^c}$. We have
 \begin{equation}\nonumber
\begin{split}
    R^\prime&=(h/m^*)^{(1/a)}=(g^c \big/\frac{g^cy^{t_j^{\prime}}}{g^by^c})^{(1/a)}\\
    &=(\frac{g^by^c}{y^{t_j^{\prime}}})^{(1/a)}\\
   &=g^{(b/a)} \cdot \frac{g^c}{g^{t_j^{\prime}}}.\\
\end{split}
\end{equation}

The successful output $g^{b/a}$ from $\mathcal{B}$ is determined by the following three events:

$\mathcal{E}_1$: No interruption is encountered during the interaction between $\mathcal{A}$ and $\mathcal{B}$.

$\mathcal{E}_2$: $\mathcal{A}$ produces a valid ciphertext $CT^*=(K^*,Z^*,R^*)$ .

$\mathcal{E}_3$: $\mathcal{E}_2$ occurs and the subscript of  $M_i^* = DSC(pk,h,CT^*,sk^*)$ is $i=j$. Then

$Pr[\mathcal{E}_1]=(1-\frac{1}{q_H})^{q_H}$,

$Pr[\mathcal{E}_2|\mathcal{E}_1]=\epsilon(\mathcal{K})$,

$Pr[\mathcal{E}_3|\mathcal{E}_1\mathcal{E}_2] =Pr[i=j|\mathcal{E}_1\mathcal{E}_2] =\frac{1}{q_H}$ .

Therefore, the advantage of $\mathcal{B}$ is

\begin{align}
\notag
Pr[\mathcal{E}_1\mathcal{E}_3]&=Pr[\mathcal{E}_1] \cdot Pr[\mathcal{E}_2|\mathcal{E}_1]\cdot Pr[\mathcal{E}_3|\mathcal{E}_1\mathcal{E}_2]\\
\notag
&=(1-\frac{1}{q_H})^{q_H}\cdot \frac{1}{q_H}\cdot \epsilon(\mathcal{K})\\
\notag
&\approx \frac{\epsilon(\mathcal{K}) }{e\cdot q_H}.
\end{align}

Since the DCDH assumption holds on $\mathbb{G}$, the advantage $\epsilon^\prime \approx \frac{\epsilon}{e\cdot q_H}$ of polytime adversary $\mathcal{B}$ is negligible. Therefore, the chameleon signcryption mechanism is SC-EUF-CMA. (Theorem 1 is proved.)

\subsection{C-EUF-CMA} \label{sec:ProofCEuf}

\textbf{Theorem 2.} If the DCDH assumption holds on $\mathbb{G}$, the proposed signcryption mechanize satisfies C-EUF-CMA.

\textbf{Proof 2.}  For contradiction, if there exists a polytime adversary $\mathcal{A}$ who wins the game $Exp_{CH\text{-}SC,\mathcal{A}}^{C\text{-}EUF}$  as shown in Fig.\ref{fig:C-EUF-CMA} with a non-negligible advantage $\epsilon$, then there exists an adversary $\mathcal{B}$ who can solve the DCDH problem with at least advantage $\epsilon^\prime$.

\begin{figure}[htbp]
  \begin{center}
        \fbox{\parbox{8.5cm}{
        \begin{flushleft}
          \hspace{0.5em}$\textbf{Experiment} \quad Exp_{CH\text{-}SC,\mathcal{A}}^{C\text{-}EUF}(\mathcal{K})$\\
          \hspace{0.5em}$Parms \leftarrow Setup(\mathcal{K});$\\
          \hspace{0.5em}$(pk,sk) \leftarrow KeyGen(Parms)$;\\
          \hspace{0.5em}$L_1\leftarrow \varnothing,L_2\leftarrow \varnothing,L_3\leftarrow \varnothing, L_s \leftarrow \varnothing, c \stackrel{R}{\leftarrow} \mathbb{Z}_q;$\\
          \hspace{0.5em}$(h,R) \leftarrow Hash(pk,M), L_1 \leftarrow (M,t,m);$\\
          \hspace{0.5em}$(CT^*,y^*) \leftarrow \mathcal{A}^{\mathcal{H}_{i\in\{1,2,3\}}(\cdot),\mathcal{SC}(\cdot)}(pk,h) $, where\\
          \hspace{2.0em}$m^\prime \leftarrow \mathcal{H}_1(M^\prime), \; L_1 \leftarrow (M^\prime,t^\prime,m^\prime)$; \\
          \hspace{2.0em}$w^\prime \leftarrow \mathcal{H}_2(K^\prime,y^\prime,Y^\prime), \; L_2\leftarrow(K^\prime,y^\prime,Y^\prime, w^\prime)  $; \\           
          \hspace{2.0em}$M^{\prime\prime} \leftarrow \mathcal{H}_3(K^\prime,Z^\prime,y^\prime), \; L_3\leftarrow(K^\prime,Z^\prime, y^\prime, M^{\prime\prime})  $; \\
          \hspace{2.0em}$CT^\prime \leftarrow \mathcal{SC}(M^\prime,y^\prime), \; L_s\leftarrow (K^\prime,Z^\prime,R^{\prime\prime})=CT^\prime$; \\
          \hspace{0.5em}if $VC(pk,h,CT^*,M,R,y^*)=1\bigwedge CT^*\notin L_s$, return 1;\\
          \hspace{0.5em}else  return 0. \\
   \end{flushleft}
        }}

    \caption{The game model of  C-EUF-CMA.}
    \label{fig:C-EUF-CMA}
    \end{center}
\end{figure}

The attack process of forging ciphertext is similar to forging signcryption. The difference is that the forged ciphertext doesn’t need to be associated with a plaintext. Details are as follows:

1) Initialization. The same as $Exp_{CH\text{-}SC,\mathcal{A}}^{SC\text{-}EUF}(\mathcal{K})$.

2) Inquiry. Before inquiring, $\mathcal{B}$ assumes that $\mathcal{A}$ will forge a signcryption on the $j$th result of inquiring $\mathcal{H}_3$. 

——$\mathcal{H}_1(M^\prime) \rightarrow m^\prime$.  $\mathcal{B}$ returns $m^\prime \leftarrow L_1(M^\prime)$ upon receiving a query on $\mathcal{H}_1$ with $M^\prime$.  If there is no corresponding entry,  $\mathcal{B}$  randomly selects $t^\prime\stackrel{R}{\leftarrow} \mathbb{Z}_q$, calculates $m^\prime = \frac{g^cy^{t^\prime}} {y^c}$, and adds a new entry $(M^\prime,t^\prime,m^\prime)$ to $L_1$. 

——$\mathcal{H}_2(K^\prime,y^\prime,Y^\prime)\rightarrow w^{\prime}$. The same as $Exp_{CH\text{-}SC,\mathcal{A}}^{SC\text{-}EUF}(\mathcal{K})$.

——$\mathcal{H}_3(K^\prime,Z^\prime, y^\prime)\rightarrow M^{\prime\prime}$. $\mathcal{B}$ returns $M^{\prime\prime} \leftarrow L_3(K^\prime,Z^\prime, y^\prime)$  upon receiving a query on $\mathcal{H}_3$ with $(K^\prime,Z^\prime, y^\prime)$. If there is no corresponding entry, $\mathcal{B}$ randomly selects $M^{\prime\prime} = M_i^{\prime\prime} \stackrel{R}{\leftarrow}\{0,1\}^n$, and adds a new entry $( K^\prime, Z^\prime, y^\prime, M^{\prime\prime})$ to $L_3$. It is worth noting that

$\cdot$ If $i=j$ and $m^{\prime\prime} \leftarrow L_1(M_{j}^{\prime\prime})$ is empty, then $\mathcal{B}$ add a new entry $(M^{\prime\prime},t^{\prime\prime} ,m^{\prime\prime})$ to $L_1$, where  $t^{\prime\prime} = t_j^{\prime\prime} \stackrel{R}{\leftarrow} \mathbb{Z}_q$ and $m^*= m^{\prime\prime} = \frac{g^cy^{t_j^{\prime\prime}}}{g^by^c}$.

$\cdot$ If $i\neq j$ and $L_1(M_{j}^{\prime\prime})$ is not empty, then $\mathcal{B}$ stops and returns ``$\perp$'' indicating an error.

——$\mathcal{SC}(M^\prime, pk^\prime) \rightarrow CT^\prime$. Upon receiving a query on $\mathcal{SC}$ with $(M^\prime,pk^\prime=y^\prime)$,  $\mathcal{B}$ returns $CT^\prime=(K^\prime,Z^\prime,R^{\prime\prime} )$  as

$k^\prime \stackrel{R}{\leftarrow} \mathbb{Z}_q$,  $K^\prime=g^{k^\prime}$,  $m^\prime \leftarrow \mathcal{H}_1(M^\prime) $, $t^\prime \leftarrow L_1(M^\prime)$,  $R^\prime=g^c/g^{t^\prime} $, $w^\prime \leftarrow \mathcal{H}_2(K,y^\prime,(y^\prime)^{k^\prime})$,  $Z^\prime=(M^\prime||R^\prime)\oplus w^ \prime$,  $M^{\prime\prime}=M_i^{\prime\prime}\leftarrow \mathcal{H}_3(K^\prime,Z^\prime,y^\prime)$, $m^{\prime\prime}\leftarrow \mathcal{H}_1(M^{\prime\prime})$, $t^{\prime\prime}\leftarrow L_1(M^ {\prime\prime})$,

$\cdot$ If $i\neq j$, then $R^{\prime\prime}=\frac{g^c}{g^{t^{\prime\prime}}}$;

$\cdot$ If $i=j$, then $\mathcal{B}$  stops and returns ``$\perp$'' indicating an error.

3) Forgery. $\mathcal{A}$ outputs $CT^*=(K^*,Z^*,R^*)$ and $y^*$. If $VC(pk,CT^*,h,M,R,y^*)=1 $ and $ CT^*\notin L_s$, then $\mathcal{A}$ wins.

In the simulation $Exp_{CH\text{-}SC,\mathcal{A}}^{C\text{-}EUF}(\mathcal{K})$,  if the guess \emph{j} from $\mathcal{B}$ is correct and $\mathcal{A}$ outputs a correct forgery,  $\mathcal{B}$  is able to output the solution $g^{b/a}=R^*\cdot\frac{g^c}{g^{t_j^{\prime\prime}}}$ of the DCDH problem based on $CT^*=(K^*,Z^*,R^*)$ , where  $M_j^{\prime\prime} \leftarrow L_3(K^*,Z^*,y^*)$, $t_j^{\prime\prime}\leftarrow L_1(M_j^{\prime\prime})$ and
\begin{equation}\nonumber
\begin{split}
    R^*&=(h/m^*)^{(1/a)}=(g^c \big/\frac{g^cy^{t_j^{\prime\prime}}}{g^by^c})^{(1/a)}\\
    &=g^{(b/a)} \cdot \frac{g^c}{g^{t_j^{\prime\prime}}}\\
\end{split}
\end{equation}

Similar to $Exp_{CH\text{-}SC,\mathcal{A}}^{C\text{-}EUF}(\mathcal{K})$, the advantage of $\mathcal{B}$ is 
\begin{align}
\notag
Pr[\mathcal{E}_1\mathcal{E}_3]&=Pr[\mathcal{E}_1] \cdot Pr[\mathcal{E}_2|\mathcal{E}_1]\cdot Pr[\mathcal{E}_3|\mathcal{E}_1\mathcal{E}_2]\\
\notag
&=(1-\frac{1}{q_H})^{2q_H}\cdot \frac{1}{q_H}\cdot \epsilon(\mathcal{K})\\
\notag
&\approx \frac{\epsilon }{e^2\cdot q_H}.
\end{align}

Since the DCDH assumption holds on $\mathbb{G}$, the advantage $\epsilon^\prime \approx \frac{\epsilon}{(e^2\cdot q_H)}$ of polytime adversary $\mathcal{B}$ is negligible. Therefore, the chameleon signcrypt algorithm is C-EUF-CMA. (Theorem 2 is proved.)

\vspace{-1.1em}

\subsection{SC-IND-CCA} \label{sec:ProofCCA}

\textbf{Theorem 3.} If the CDH assumption holds on $\mathbb{G}$, the proposed signcryption mechanism satisfies SC-IND-CCA.

\textbf{Proof 3:}  For contradiction, if there is a polytime adversary $\mathcal{A}$ who wins the game $Exp_{CH\text{-}SC,\mathcal{A}}^{IND\text{-}CCA}$ as shown in Fig.\ref{fig:SC-IND-CCA} with a non-negligible advantage $\epsilon$, there is an adversary $\mathcal{B}$ who can solve the CDH problem with at least advantage $\epsilon^\prime$.

\begin{figure}[htbp]
  \begin{center}
        \fbox{\parbox{8.5cm}{
        \begin{flushleft}
          \hspace{0.5em}$\textbf{Experiment} \quad Exp_{CH\text{-}SC,\mathcal{A}}^{IND\text{-}CCA}(\mathcal{K})$\\
          \hspace{0.5em}$parm \leftarrow Setup(\mathcal{K});$\\
          \hspace{0.5em}$(pk_U,sk_U) \leftarrow KeyGen(parm)$;\\
          \hspace{0.5em}$L_1\leftarrow \varnothing,L_2\leftarrow \varnothing,L_3\leftarrow \varnothing,L_s \leftarrow \varnothing, c \stackrel{R}{\leftarrow} \mathbb{Z}_q;$\\
          \hspace{0.5em}$(h,R) \leftarrow Hash(pk,M), L_1 \leftarrow (M,t,m);$\\
          \hspace{0.5em}$(m_0,m_1) \leftarrow \mathcal{A}^{\mathcal{H}_{i\in \{1,2,3\}}(\cdot),\mathcal{SC}(\cdot),\mathcal{DSC}(\cdot)}(pk,h)$, \\
          \hspace{1.5em}$m^\prime \leftarrow \mathcal{H}_1(M^\prime), \; L_1 \leftarrow (M^\prime,t^\prime,m^\prime)$; \\
          \hspace{1.5em}$M^{\prime\prime} \leftarrow \mathcal{H}_3(K^\prime,Z^\prime, y^\prime), \; L_3\leftarrow(K^\prime,Z^\prime, y^\prime M^{\prime\prime})  $; \\
          \hspace{1.5em}$w^\prime \leftarrow \mathcal{H}_2(K^\prime,y_R,Y^\prime), \; L_2\leftarrow(K^\prime,y_R,Y^\prime, w^\prime)  $; \\
          \hspace{1.5em}$CT^\prime \leftarrow \mathcal{SC}(pk_U,M^\prime,pk_R), \; L_s\leftarrow (M^\prime,CT^\prime)$; \\
          \hspace{1.5em}$M^\prime \leftarrow \mathcal{DSC}(pk_S,CT^\prime, pk_U), \; L_2\leftarrow (K^\prime,y_U,\perp,w^\prime)$; \\
          \hspace{0.5em}$b\stackrel{R}{\leftarrow} \{0,1\}$,$CT^* \leftarrow \mathcal{SC}(pk_U,M_b,pk_R)$;\\
          \hspace{0.5em}$\hat{b} \leftarrow \mathcal{A}^{\mathcal{H}_{i\in \{1,2,3\}}(\cdot),\mathcal{SC}(\cdot),\mathcal{DSC}(\cdot)}(pk,h)$;\\
          \hspace{0.5em}if $\hat{b}=b$, return 1; else return 0. \\
        \end{flushleft}
        }}

    \caption{The game model of  SC-IND-CCA.}
    \label{fig:SC-IND-CCA}
    \vspace{-1.1em}
    \end{center}
\end{figure}

Suppose $\mathcal{B}$ is given a random instance  $(g, g^a, g^b) \in \mathbb{G}^3$ of the CDH problem. Based on the instance, $\mathcal{B}$ sets up a simulated environment of SC-IND-CCA as $Exp_{CH\text{-}SC,\mathcal{A}}^{IND\text{-}CCA}(\mathcal{K})$ for $\mathcal{A}$ and runs $\mathcal{A}$ as a subroutine to ﬁnd the solution $g^{ab}$.

1) Initialization. The challenger $\mathcal{B}$ sets the public key  $pk_U=y_U=g^u$ and initializes the chameleon tuple $(h, M, R)$ and the lists $L_1,L_2,L_3,L_s$ for oracles $\mathcal {H}_1,\mathcal{H}_2,\mathcal{H}_3$, $\mathcal{SC}$ and $\mathcal{DSC}$. Among them,  $L_1=(M^\prime,t^\prime,m^\prime)$, $L_2=(K^\prime,Z^\prime,M^{\prime\prime})$, $L_3=(K^\prime,y^\prime,Y^\prime,w^\prime)$, $L_s=(M^\prime, K^\prime,Z^\prime,R^{\prime\prime},pk^\prime)$. Besides, $\mathcal{B}$ adds a entry $ (M,t,m)$ to $L_1$, where $c\stackrel{R}{\leftarrow} \mathbb{Z}_q$, $h=g^c$, $t\stackrel{R}{\leftarrow} \mathbb{Z}_q$, $m=\frac{ g^cy^t}{y^c}$, $R=\frac{g^c}{g^t}$.

2) Inquiry 1. $\mathcal{A}$ makes a certain number of inquiries on hash oracles $\mathcal{H}_{i\in \{1,2,3\}}$, the signcryption oracle $\mathcal{SC}(pk_U,M,pk_R) $, and the de-signcryption oracle $\mathcal{DSC}(pk_S,CT,pk_U)$  as follows:

——$\mathcal{H}_1(M^\prime) \rightarrow m^\prime$.  $\mathcal{B}$ returns $m^\prime \leftarrow L_1(M^\prime)$ upon receiving a query on $\mathcal{H}_1$ with $M^\prime$.  If there is no corresponding entry,  $\mathcal{B}$  randomly selects $t^\prime\stackrel{R}{\leftarrow} \mathbb{Z}_q$, calculates $m^\prime = \frac{g^cy_U^{t^\prime}} {y_U^c}$, and adds a new entry $(M^\prime,t^\prime,m^\prime)$ to $L_1$. 

——$\mathcal{H}_2(K^\prime,y_R^\prime,Y^\prime)\rightarrow w^{\prime}$. $\mathcal{B}$ returns $w^{\prime} \leftarrow L_2(K^\prime,y_R^\prime,Y^\prime)$  upon receiving a query on  $\mathcal{H}_2$ with $(K^\prime,y_R^\prime,Y^\prime)$. If there is no corresponding entry, $\mathcal{B}$ randomly selects $w^\prime\stackrel{R}{\leftarrow}\{0,1\}^{n+l}$ and adds a new entry $( K^\prime, y_R^\prime, Y^\prime, w^\prime)$ to $L_2$.

——$\mathcal{H}_3(K^\prime,Z^\prime, y_R^\prime)\rightarrow M^{\prime\prime}$. $\mathcal{B}$ returns $M^{\prime\prime} \leftarrow L_3(K^\prime,Z^\prime, y_R^\prime)$  upon receiving a query on $\mathcal{H}_3$ with $(K^\prime,Z^\prime, y_R^\prime)$. If there is no corresponding entry, $\mathcal{B}$ randomly selects $M^{\prime\prime} = M_i^{\prime\prime} \stackrel{R}{\leftarrow}\{0,1\}^n$ and adds a new entry $( K^\prime, Z^\prime, y_R^\prime, M^{\prime\prime})$ to $L_3$.

—— $\mathcal{SC}(pk_U,M^\prime,pk_R^\prime) \rightarrow CT^\prime$. Upon receiving a query on $\mathcal{SC}$ with $(pk_U,M^\prime,pk_R^\prime=y_R^\prime)$,  $\mathcal{B}$ checks $y_R^\prime$  and returns $CT^\prime$ as  

$\cdot$ If $y_U=y_R$, $\mathcal{B}$ returns ``$\perp$'';  

$\cdot$ If $y_U \neq y_R$, $\mathcal{B}$ returns $CT^\prime=(K^\prime,Z^\prime,R^{\prime\prime} )$, where  

$k^\prime \stackrel{R}{\leftarrow} \mathbb{Z}_q$,  $K^\prime=g^{k^\prime}$,  $m^\prime \leftarrow \mathcal{H}_1(M^\prime) $, $t^\prime \leftarrow L_1(M^\prime)$,  $R^\prime=g^c/g^{t^\prime} $, $Y^\prime ={(y_R^{\prime})}^{k^\prime}$,$w^\prime \leftarrow \mathcal{H}_2(K,y_R^\prime,Y^\prime)$,  $Z^\prime=(M^\prime||R^\prime)\oplus w^ \prime$,  $M^{\prime\prime}\leftarrow \mathcal{H}_3(K^\prime,Z^\prime,y_R^\prime)$, $m^{\prime\prime}\leftarrow \mathcal{H}_1(M^{\prime\prime})$, $t^{\prime\prime}\leftarrow L_1(M^ {\prime\prime})$, $R^{\prime\prime}=\frac {g^c}{g^{t^{\prime\prime}}}$.

—— $\mathcal{DSC}(pk_S^\prime, CT^\prime,pk_U) \rightarrow M^\prime $. Upon receiving a query on $\mathcal{DSC}$ with $(pk_S^\prime=y_S^\prime,CT^\prime,pk_U=y_U)$ , where $CT^\prime=(K^\prime,Z^\prime,R^{\prime\prime}$), $\mathcal{B}$ returns $M^\prime$ as

(i) $\mathcal{B}$ retrieves $M^{\prime\prime} \leftarrow \mathcal{H}_3(K^\prime,Z^\prime,y_U)$ , $m_i^{\prime\prime} \leftarrow \mathcal{H}_1(M_i ^{\prime\prime})$ and checks $e(h/m^{\prime\prime},g)\stackrel{?}{=}e(R^{\prime\prime},y_S^\prime)$. If the equation doesn't hold, $\mathcal{B}$ returns ``$\perp$''.

(ii) $\mathcal{B}$ retrieves $(K^\prime,y_U, Y_i^\prime,w_i^\prime)_{i\in [1,q_H]} \leftarrow L_3(K^\prime,y_U)$. If there is an entry satisfying $Y_i^\prime=\top$ or $e(K^\prime,y_U)=e(Y_i^\prime,g )$,  $\mathcal{B}$ sets $w^\prime=w_i^\prime$; otherwise, $\mathcal{B}$ randomly selects $w^\prime \stackrel{R}{\leftarrow}\{0,1\}^{n+l}$ and adds a new entry $(K^\prime,y_U,\top,w^\prime)$ to $L_2$,  where the symbol ``$\top$'' represents the value to be solved for the CDH instance. Finally,  $\mathcal{B}$ calculates $(M^\prime||R^\prime)= Z^\prime \oplus w^\prime$.

(iii)  $\mathcal{B}$ retrieves $m^\prime \leftarrow \mathcal{H}_1(M^{\prime})$ and checks $e(h/m^\prime,g)\stackrel{?}{=}e(R^{\prime\prime}, y_S^\prime)$. If the equation holds, $\mathcal{B}$ outputs $M^\prime$ from the decryption result $(M^\prime||R^\prime)$; otherwise,  $\mathcal{B}$ returns ``$\perp$''.

3) Challenge. $\mathcal{A}$ submits $\mathcal{B}$ with two equal-length plaintexts $m_0,m_1 \in \{0,1\}^n$, $|m_0|=|m_1|$. $\mathcal{B}$ returns a signcryption $CT^*= (K^*, Z^*, R^*)$ as

$K^*=g^a$, $b \stackrel{R}{\leftarrow}\{0,1\}$, $m^\prime\leftarrow \mathcal{H}_1(M_b)$ , $t^\prime \leftarrow L_1(M_b)$, $R^\prime=g^c/g^{t^\prime}$ , $Z^* \stackrel{R}{\leftarrow} \{0,1\}^{n+l}$,  $w^* = Z^* \oplus (M_b||R^\prime)$, $M^{\prime\prime}\leftarrow \mathcal{H}_3(K^*,Z^*,y_U) $, $t^{\prime\prime}\leftarrow L_1(M^{\prime\prime})$, $R^*=g^c/g^{t^{\prime\prime}}$, $L_2 \leftarrow (K^*,y_U,\top, w^*)$,  $L_s \leftarrow (M_b, K^*,Z^*,R^*)$.

4) Inquiry 2. $\mathcal{A}$ performs a certain number of oracle queries as inquiry 1 but restricts it from inquiring $\mathcal{DSC}$ with $CT^*$. If the query $\mathcal{H}_2(K^\prime=g^a,y_R^\prime=g^b,\lambda)$ satisfies $e(K,y_R)=e(\lambda,g)$,  $\mathcal{B}$ outputs $\lambda$ and stops.

5) Finally,  $\mathcal{A}$ outputs a guess value $\hat{b}$. If $\hat{b}=b$,  $\mathcal{A}$ wins.

To make the ideal environment $Exp_{CH\text{-}SC,\mathcal{A}}^{IND\text{-}CCA}(\mathcal{K})$ indistinguishable from the real attack, the simulations on $\mathcal{H}_1$, $\mathcal{H}_2$, $\mathcal{H}_3$, $\mathcal{SC}$ and $\mathcal{DSC}$ need to satisfy the perfect, where $\mathcal{H}_{1\in \{1,2,3\}}$and $\mathcal{SC}$ are perfect in above simulations.

What makes $\mathcal{DSC}$ imperfect is that the oracle related to $\mathcal{DSC}$ rejects the valid signcryption, specifically,  $(K,y_U, \top) \in L_2$ and $M^\prime||R^\prime=Z^\prime\oplus \mathcal{H}_2(K,y_U,\top)$. This rejection implies that $\mathcal{A}$ does not query $(K,y_U,\lambda)$ on $\mathcal{H}_2$ such that $e(K,y_U)=e(\lambda,g)$ , where $Z_2 =W_2\oplus R^\prime$, $Z^\prime=(Z_1,Z_2)$, $(W_1,W_2)=H_2(K,y_U,\top)$,
$$Pr[Z_2=W_2\oplus R^\prime] \leq \frac{q_H}{|\mathbb{G}|} = \frac{q_H}{2^k}.$$
Therefore, the probability that $\mathcal{B}$ does not fail is

$$Pr[\mathcal{B}\;\text{not fail} ] = 1-\frac{q_H}{2^k}.$$

Let $E$ indicate that $\mathcal{A}$ has asked  $\mathcal{H}_2$ with $(K^\prime=g^a,y_R^\prime=g^b, \lambda=Y^\prime=(y_R^\prime)^a)$  during the simulation process. Then, the winning probability of $\mathcal{A}$ is
\begin{equation}\nonumber
    \begin{split}
    \epsilon &= \big| Pr[\hat{b}=b]-\frac{1}{2}\big|\\
    &\leq \big|Pr[E] \cdot Pr[\hat{b}=b|{E}]+Pr[^\neg E] \cdot Pr[\hat{b}=b|^\neg E] - \frac{1}{2}\big|\\         & \leq \big| Pr[E] + \frac{1}{2}Pr[^\neg E] - \frac{1}{2} \big|    \\
    &=\frac{1}{2}Pr[E].\\
    \end{split}
\end{equation}

Where $Pr[\hat{b}=b|^\neg E]=\frac{1}{2}$ and $Pr[E]\geq 2 \epsilon$. Since the solution $g^{ab}$ output by $\mathcal{B}$ depends on $\mathcal{A}$'s queries, the probability that $\mathcal{B}$ solves the CDH problem is 
\begin{equation}\nonumber
    \begin{split}
    \epsilon^\prime &= Pr[E \wedge (\mathcal{B} \; \text{not fail})]\\
    &= Pr[E] \cdot Pr[\mathcal{B} \; \text{not fail}]\\
    &\geq 2\epsilon (1-\frac{q_H}{2^k})\\
    \end{split}
\end{equation}
Since the CDH assumption holds on $\mathbb{G}$, the advantage $\epsilon^\prime \geq 2\epsilon (1-\frac{q_H}{2^k})$ is negligible. Therefore, the chameleon signcrypt satisfies SC-IND-CCA. (Theorem 3 is proved.)

\subsection{Public Verifiability} \label{sec:ProofPubVeri}

\textbf{Theorem 4.} If the CDH assumption holds on $\mathbb{G}$ and $H_3$ satisfies collision resistance, the proposed signcryption mechanism satisfies public verifiability.

To violate the public verifiability, the adversary $\mathcal{C}$'s goal is to construct $pk_C$ and $(h_C,\tilde{M}_C,\tilde{R}_C)$ such that $VC (pk_C$ , $CT_{AB}$, $h_C$, $\tilde{M}_C, \tilde{R}_C, pk_B )=1$ or $VC (pk_A, CT_{AB}$, $h_C$, $\tilde{M}_C$, $\tilde{R}_C$, $pk_C )=1$ .

Case 1: To make $VC (pk_C, CT_{AB}, h_C,\tilde{M}_C,\tilde{R}_C, pk_B )=1$, the related parameters need to satisfy $Check(pk_C$, $h_C$, $\tilde{M}_C$, $\tilde{R}_C)=1$ and $Check(pk_C$, $h_C,H_3(K,Z,pk_B), R^{\prime\prime})=1$. Since $\mathcal{C}$ has the corresponding private key $sk_C$, he/she can construct any $(h_C,\tilde{M}_C,\tilde{R}_C)$  satisfying $Check(pk_C$, $h_C,\tilde{ M}_C,\tilde{R}_C)=1$. Since $(K,Z,R^{\prime\prime})=CT_{AB}$ and $(pk_C$, $pk_B)$ are fixed, $\mathcal{C}$ only needs to construct $h_C$ satisfying $Check(pk_C$, $h_C$, $H_3(K,Z,pk_B),R^{\prime\prime})=1$, that is, $e(h_C/H_3(K,Z,pk_B),g)=e(R^{\prime\prime},pk_C)$. Let $m=H_3(K,Z,pk_B)$. Then, $(g,R^{\prime\prime},pk_C)=(g,g^a,g^b)\in \mathbb{G}^3$ is a given CDH problem instance. If the $h_C$ constructed by $\mathcal{C}$ satisfies $e(h_C/m,g)=e(g^a,g^b)$ with a non-negligible advantage, then the adversary can output the solution $g^{ab}=h_C/H_3(K,Z,pk_B)$ with the equal advantage, which contradicts the CDH assumption.

Case 2: To make $VC (pk_A, CT_{AB}, h_C,\tilde{M}_C,\tilde{R}_C, pk_C )=1$, the related parameters must satisfy $Check(pk_A$, $h_C$, $\tilde{M}_C$, $\tilde{R}_C)=1$ and $Check(pk_A,h_C, H_3(K,Z,pk_C)$,$R^{\prime\prime})=1$. Since $(h_A,\tilde{M}_A,\tilde{R}_A)$ is a public chameleon triplet of $pk_A$, the parameters $(h_C,\tilde{M}_C,\tilde{R}_C)=(h_A,\tilde{M}_A,\tilde{R}_A)$ satisfy $Check(pk_A$, $h_C,\tilde{M }_C,\tilde{R}_C)=1$. To satisfy $Check(pk_A$, $h_C$, $H_3(K$, $Z$, $pk_C)$, $R^{\prime\prime})=Check(pk_A$, $h_A$, $H_3(K,Z,pk_C),R^{\prime\prime})=1$, it is necessary to construct a public key $pk_C$ such that $H_3(K,Z,pk_C)=H_3 (K,Z,pk_B)$ and $pk_C \neq pk_B$, which contradicts the collision resistance of $H_3$. 

To sum up, in our signcryption mechanism, anyone can verify the identities $A$ and $B$ implying in $CT_{AB}$ through $pk_A$ and $pk_B$ without decryption. The adversary cannot complete identity verification based on the forged $pk_C$ and parameters $(h_C,\tilde{M}_C,\tilde{R}_C)$. Therefore, the signcryption mechanism satisfies public verifiability.

\begin{IEEEbiography}[{\includegraphics[width=1in,height=1.25in,clip,keepaspectratio]{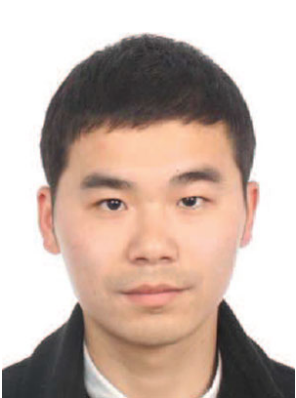}}]{Zhenyong Zhang}
 (Member, IEEE) received the achelor’s degree from Central South University, Changsha, China, in 2015, and the Ph.D. degree from Zhejiang University, Hangzhou, China, in 2020. He was a Visiting Scholar with Singapore University of Technology and Design, Singapore, from 2018 to 2019.  He is currently a Professor with the College of Computer Science and Technology, Guizhou University, Guiyang, China. His research interests include cyber–physical system security, applied cryptography, metaverse security, and machine learning security.
\end{IEEEbiography}

\begin{IEEEbiography}[{\includegraphics[width=1in,height=1.25in,clip,keepaspectratio]{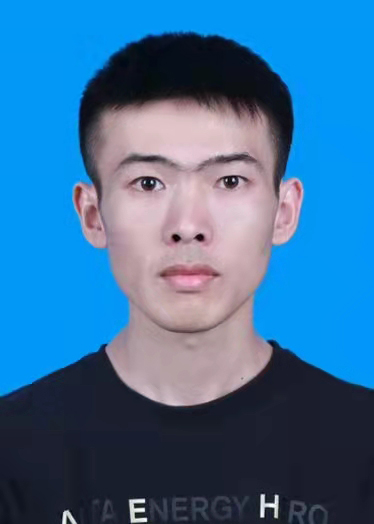}}]{Kedi Yang}
received the B.Sc. degree in mathematics and applied mathematics from Anshun University in 2012, and the M.Sc. degree in applied mathematics from Guizhou University in 2020.
He is currently a Ph.D candidate in the College of Computer Science and Technology, Guizhou University, Guiyang, China. His research interests mainly focus on Metaverse security, data provenance, and blockchain technology.
\end{IEEEbiography}

\begin{IEEEbiography}[{\includegraphics[width=1in,height=1.25in,clip,keepaspectratio]{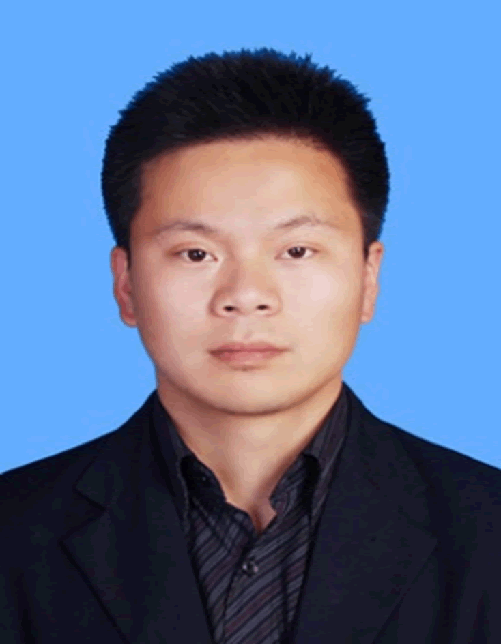}}]{Youliang Tian}
 (Member, IEEE) received the B.Sc. degree in mathematics and applied mathematics and the M.Sc. degree in applied mathematics from Guizhou University, in 2004 and 2009, respectively, and the Ph.D. degree in cryptography from Xidian University, in 2012. From 2012 to 2015, he was a Postdoctoral Associate with the State Key Laboratory for Chinese Academy of Sciences. He is currently a Professor and a Ph.D. Supervisor with the College of Computer Science and Technology, Guizhou University. His research interests include algorithm game theory, cryptography, and security protocol.
\end{IEEEbiography}

\begin{IEEEbiography}[{\includegraphics[width=1in,height=1.25in,clip,keepaspectratio]{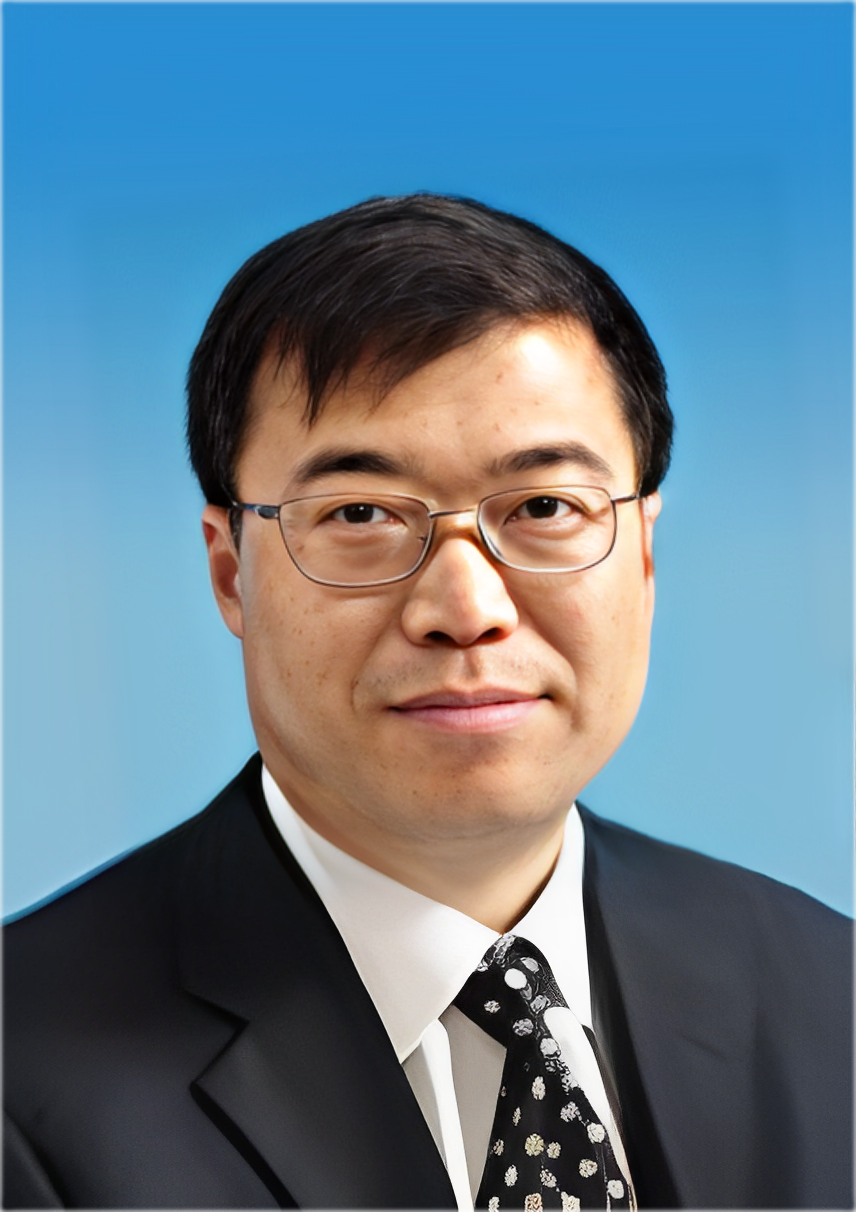}}]{Jianfeng Ma}
 (Member, IEEE) received the B.S. degree in mathematics from Shaanxi Normal University, Xi'an, China, in 1985, and the M.S. degree and the Ph.D. degree in computer software and telecommunication engineering from Xidian University, Xi'an, China, in 1988 and 1995, respectively. He is currently a professor with the School of Cyber Engineering, Xidian University, Xi'an, China. He is also the Director of the Shaanxi Key Laboratory of Network and System Security. His current research interests include information and network security and mobile computing systems.
\end{IEEEbiography}

\end{document}